\documentclass{aa}

\usepackage{booktabs}
\usepackage{graphicx}
\usepackage[english]{babel}
\usepackage{multirow}
\usepackage{xcolor}
\usepackage{graphicx}
\usepackage{natbib} 
\bibpunct{(}{)}{;}{a}{}{,} 
\usepackage{gensymb}
\usepackage{ftnxtra}
\usepackage{fnpos}
\usepackage{txfonts}
\usepackage{soul}
\setstcolor{red}

\begin{document}

   \title{Absolute dimensions and apsidal motion of the eclipsing binaries V889 Aql and V402 Lac}

   \author{D.~Baroch\inst{1,2} \and A.~Giménez\inst{3} \and J.~C.~Morales\inst{1,2} \and I.~Ribas\inst{1,2} \and E.~Herrero\inst{1,2} \and V.~Perdelwitz\inst{4,5} \and C.~Jordi\inst{6,2} \and T.~Granzer\inst{7} \and C.~Allende Prieto\inst{8,9}
          }
   \authorrunning{D.~Baroch et al.}
   \titlerunning{Absolute dimensions and apsidal motion of the eclipsing binaries V889 Aql and V402 Lac}
   \institute{Institut de Ci\`encies de l'Espai (ICE, CSIC),
              Campus UAB, c/~Can Magrans s/n, E-08193 Bellaterra, Barcelona, Spain\\
              \email{baroch@ieec.cat}
         \and
              Institut d'Estudis Espacials de Catalunya (IEEC),
              c/ Gran Capit\`a 2-4, E-08034 Barcelona, Spain
         \and
              Centro de Astrobiolog\'{\i}a (CSIC-INTA), Ctra. Ajalvir, km 4. E-28850 Torrej\'on de Ardoz, Madrid, Spain
        \and
            Department of Physics, Ariel University, 40700 Ariel, Israel
        \and
            Hamburger Sternwarte, Universit\"at Hamburg, Gojenbergsweg 112, D-21029 Hamburg, Germany
        \and
            Dept. Física Quàntica i Astrofísica, Institut de Ciències del Cosmos (ICCUB), Universitat de Barcelona (IEEC-UB), Martí i Franquès 1, E-08028 Barcelona, Spain
        \and
            Leibniz-Institut f\"ur Astrophysik Potsdam (AIP), An der Sternwarte 16, D-14482 Potsdam, Germany
        \and
            Instituto de Astrofísica de Canarias, Vía Láctea S/N, E-38205 La Laguna, Tenerife, Spain
        \and
            Universidad de La Laguna, Departamento de Astrofísica, E-38206 La Laguna, Tenerife, Spain
              }

   \date{Received 16 June 2022 / Accepted 24 June 2022} 

 
  \abstract
   {Double-lined eclipsing binaries allow the direct determination of masses and radii, which are key to test stellar models. With the launch of the {\em TESS} mission, many well-known eclipsing binaries have been observed at higher photometric precision, permitting the improvement of the absolute dimensions determinations.}
   {Using {\em TESS} data and newly-obtained spectroscopic observations, we aim at determining the masses and radii of the eccentric eclipsing binary systems V889\,Aql and V402\,Lac, together with their apsidal motion parameters.}
   {We modelled simultaneously radial velocity curves and times of eclipse for each target to precisely determine the orbital parameters of the systems, which we used to analyse the light curves and then obtain their absolute dimensions. We compared the obtained values with those predicted by theoretical models.}
   {We determined masses and radii of the components of both systems with relative uncertainties lower than 2\,\%. V889\,Aql is composed of two stars with masses $2.17\pm0.02$\,M$_{\odot}$ and $2.13\pm0.01$\,M$_{\odot}$ and radii $1.87\pm0.04$\,R$_{\odot}$ and $1.85\pm0.04$\,R$_{\odot}$. We found conclusive evidence of the presence of a third body orbiting V889\,Aql with a period of 67\,years. Based on the detected third light and the absence of signal in the spectra, we suggest that this third body could in turn be a binary composed by two $\sim$1.4\,M$_{\odot}$ stars. V402\,Lac is composed by two stars with masses $2.80\pm0.05$\,M$_{\odot}$ and $2.78\pm0.05$\,M$_{\odot}$ and radii $2.38\pm0.03$\,R$_{\odot}$ and $2.36\pm0.03$\,R$_{\odot}$. The times of minimum light are compatible with the presence of a third body for this system too, although its period is not yet fully sampled. In both cases we have found a good agreement between the observed apsidal motion rates and the model predictions.}
   {}
 

   \maketitle

%
\section{Introduction}

Well-detached double-lined eclipsing binaries are the basic source of information for stellar dimensions, such as masses and radii, to high precision \citep{Andersen1991,Torres2010}. Moreover, eccentric systems showing apsidal motion, i.e., the precession of the line of apsides, yields additional unique information about the stellar internal density distribution through the $\log k_2$ parameter. This can be only accurately determined for this type of eclipsing binaries, where the masses and radii of their components can be measured to a precision better than $\sim2$\% \citep{Claret2021}.

Observed apsidal motion rates have two main components: a secular variation due to the non-spherical shape of the stars, called the quadrupole or classical term, and another contribution due to the effect in the two-body equations of motion given by General Relativity, the relativistic term. Both components have recently been studied in \cite{Baroch2021} and \cite{Claret2021} for those systems with precise times of minimum light determinations based on photometric measurements obtained with the {\em TESS} space mission \citep{Ricker2015}, which we also use in this paper. An additional element of the apsidal motion rate can be caused by the presence of a perturbing third body \cite{Borkovits2007,Borkovits2019}. Although this term is generally negligible when compared to the other terms, it may become relevant in precise determinations of the apsidal motion rate or when the classical term is relatively small.

Two well-known cases of eclipsing binaries showing apsidal motion and with {\em TESS} data were not included in \cite{Baroch2021} and \cite{Claret2021} because of the lack of sufficiently precise  absolute dimensions. These are V889\,Aql and V402\,Lac. Both systems show quite eccentric orbits, with $e\sim0.4$, but different orbital periods. This leads to different dominant terms of the apsidal motion: relativistic for V889\,Aql and classical for V402\,Lac. 

The system V889\,Aql was identified 4 decades ago as the first case of relativistic apsidal motion \citep{Gimenez1982}, based on the comparison of the displacement of the secondary eclipse compared with a previous light curve by \cite{Semeniuk1967}. A value of $\dot{\omega}$ = 0.00046$\pm$0.00015\,deg\,cycle$^{-1}$ was obtained adopting an orbital eccentricity of 0.37 deduced from the light curve analysis. \cite{Khaliullin1989} obtained a new light curve and confirmed the apsidal motion rate given by \cite{Gimenez1982}. Nevertheless, a deeper analysis of the light curve yielded a refined determination of the orbital eccentricity, the presence of significant third light ($L_3=$18.5\% of the total light) as well as indications of the presence of a third body. This was later confirmed, with more precise times of minimum light, by \cite{Wolf2005} establishing a period for the third body orbit of 52 years. Nevertheless, the estimated mass of the third body, assuming a co-planar orbit, could not explain the observed $L_{3}$ value. In any case, the comparison of the observed apsidal motion rate with theoretical predictions requires a good radial velocity curve, allowing a precise determination of the stellar masses and eccentricity. A more recent study by \cite{Kiran2019} included spectroscopic measurements but could not achieve the required precision. Moreover, the authors used an incomplete light curve, with no consideration of $L_3$. A new, more precise light curve, with no phase gaps, as well as a better radial velocity curve, are clearly needed.

The system V402\,Lac was studied using Hipparcos photometry by \cite{Bulut2008}, who obtained initial results with the EBOP light curve analysis tool. \cite{Herrero2010} presented new photometry and analysed the available radial velocity curves but the degeneracy between eccentricity, flux ratio and ratio of radii could not be solved and the presence of third light was not considered. New photometry by \cite{Hoyman2018} produced an additional light curve, and analysed the times of eclipse available, leading to an apsidal motion period of 213 years and the detection of a third body, with an orbital period of 20.5 years around the eclipsing binary. Unfortunately a constraint of $L_2/L_1$ from spectroscopy could not be established and the presence of third light in the crowded field of view was not analysed. To address these issues, {\em TESS} measurements provide additional times of minima and a precise light curve while new spectroscopic observations allowed constraining the possible solutions.

In this work, new spectroscopic measurements have been obtained for both stars and their {\em TESS} light curves, as well as new eclipse timings, have been analysed. As a result, we present new absolute dimensions and orbital parameters. We show the general properties and sidereal orbital periods ($P_{\rm s}$) of the two objects in Table\,\ref{tab:proprties}.

\begin{table}[!t]
\centering
\caption{General properties of V889\,Aql and V402\,Lac.}
\label{tab:proprties}
\begin{tabular}{l ccc} 
\hline\hline
\noalign{\smallskip}
\multirow{2}{*}{Parameter} & \multicolumn{2}{c}{Star} & \multirow{2}{*}{Ref.}  \\
\cline{2-3}
\noalign{\smallskip}
& V889\,Aql & V402\,Lac  & \\
\noalign{\smallskip}
\hline
\noalign{\smallskip}
$\alpha$ (J2016) & 19:18:49.84 & 22:09:15.20 & (1)  \\
$\delta$ (J2016) & +16:15:00.4 & +44:50:47.3 & (1)  \\
$\varpi$ [mas] & 2.96$\pm$0.02 & 3.54$\pm$0.09 & (1) \\
$d$ [pc] & 338$\pm$2 & 283$\pm$8 & (1)  \\
$G$ [mag] & 8.532$\pm$0.003 & 6.723$\pm$0.003 & (1) \\
Sp.\,T & B9.5\,V & B8\,V &  (2, 3) \\
$P_{\rm s}$ [d] & 11.120757$\pm$0.000004 & 3.782043$\pm$0.000005 & (4) \\
\noalign{\smallskip}
\hline
\end{tabular}
\tablebib{ 
(1): \cite{Gaia2020}; (2): \cite{Abt2004}; (3): \cite{Wolf2010}; (4): This work.
}
\end{table}

Section\,\ref{sec:obs} presents the observational data, both from spectroscopy and photometry. Section\,\ref{sec:orbital} describes the analysis of the radial velocity curves and the times of eclipse, leading to the orbital parameters. Section\,\ref{sec:dimensions} presents the modelling of the {\em TESS} light curves, which yield the corresponding absolute dimensions of the system. Finally, Section\,\ref{sec:conclusions} compares the results with theoretical predictions and summarises the conclusions.

\section{Observations}
\label{sec:obs}
\subsection{Spectroscopy}

We obtained new spectroscopic observations of V899\,Aql and V402\,Lac with four different spectrographs. Observations of the two targets between September and October 2020 were acquired with CAFE \citep{Aceituno2013}, a high resolution échelle spectrograph mounted on the 2.2\,m telescope at Calar Alto observatory, Spain, which covers a wavelength range between 365\,nm and 980\,nm with a spectral resolution of 70\,000. V889\,Aql was also observed between March and April 2021 with the échelle spectrograph mounted on the 1.2\,m telescope TIGRE \citep{Schmitt2014} at la Luz observatory, Mexico. The spectrograph at TIGRE has a spectral resolution of 20\,000 and covers a wavelength range between 350\,nm and 880\,nm in two channels, the blue channel between 350\,nm and 560\,nm, and the red channel between 580\,nm and 880\,nm. Here we only use data from the blue channel of TIGRE, since it is closer to the emission peak of V889\,Aql. 

Additionally, we also observed V402\,Lac between October 2001 and August 2006, and between September and November 2009 with the Robert G. Tull Coudé Spectrograph \citep{Tull1995} and with the STELLA échelle spectrograph \citep{Strassmeier2001}, respectively. The Tull spectrograph is mounted on the 2.7\,m telescope at the McDonald Observatory, USA, and covers a wavelength range between 375\,nm and 1020\,nm at a resolving power of about 60\,000. The STELLA échelle spectrograph is mounted on the two 1.2\,m telescopes at Izana Observatory, Spain, and covers a wavelength range between 390\,nm and 870\,nm with a spectral resolution of about 55\,000. Data from these two instruments were already used in the analysis of \cite{Herrero2010}.  We summarise the number of available spectra from each instrument in Table\,\ref{tab:specdata}, together with the time span of each dataset.

\begin{table}[!t]
\centering
\caption{Number and time span of the available spectra for each target coming from different instruments, with the number of RVs used in the analysis in parenthesis.}
\label{tab:specdata}
\begin{tabular}{l cccc} 
\hline\hline
\noalign{\smallskip}
& \multicolumn{2}{c}{V889\,Aql} & \multicolumn{2}{c}{V402\,Lac} \\
\cline{2-5}
\noalign{\smallskip}
& N(N$_{\rm used}$) & $\Delta t$ [d] & N(N$_{\rm used}$) & $\Delta t$ [d] \\
\noalign{\smallskip}
\hline
\noalign{\smallskip}
CAFE & 27(19) & 46 & 39(27) & 47 \\
TIGRE & 15(15) & 29 & $\cdots$ & $\cdots$ \\
STELLA & $\cdots$ & $\cdots$ & 48(38) & 79 \\
Tull & $\cdots$ & $\cdots$ & 33(30) & 1771\\
Total & 42(34) & 227 & 120(95) & 6944 \\
\noalign{\smallskip}
\hline
\end{tabular}
\end{table}

The radial velocities (hereafter RVs) of the components of each system were computed by searching for their signature in the spectra using \texttt{todmor} \citep{Zucker2003}, an implementation for multi-order spectra of the two-dimensional cross-correlation method \texttt{todcor} \citep{Zucker1994}. In this technique, two different template spectra are used for each component, scaled according to their flux ratio in the wavelength range of the observations. The two spectra are simultaneously Doppler-shifted, constructing a two-dimensional cross-correlation function (CCF) map from which the RVs of the two components are computed. As templates for the calculation of the CCFs, we employed synthetic Phoenix stellar models \citep{Husser2013} for the observations of V889\,Aql. However, since the expected temperature of V402\,Lac is close to the limit of the Phoenix models, 12\,000\,K \citep{Herrero2010,Hoyman2018}, we instead decided to use models from the Coelho synthetic stellar library \citep{Coelho2014}, which covers temperatures up to 25\,000\,K, and allowed us to explore a wider grid of template temperatures. We adopted solar metallicities for all the spectra. For each target and instrument, we checked all the \'echelle orders of the spectra with the highest signal-to-noise ratio (SNR) with \texttt{todmor}, and discarded those with no signals in their CCFs due to telluric contamination or lack of spectral lines.

The parameters of the best templates were found by exploring a grid of values for effective temperatures ($T_{\rm eff}$), luminosity ratios ($L_2/L_1$), and spectral line broadening ($v_{\rm b}$) due to rotation, spectral resolution, and other effects such as macroturbulence, seeking the combination that maximised the CCF peak of each spectrum. In the case of V889\,Aql, a luminosity ratio of $L_2/L_1=0.98\pm0.06$ was derived, together with the $T_{\rm eff}$ and $v_{\rm b}$ values for each instrument given in Table\,\ref{tab:todmorV889}, and with the weighted mean values listed in the last row. In the case of V402\,Lac, however, we found strong degeneracies between $L_2/L_1$ and $v_{\rm b}$, probably due to the lack of clear and deep lines in the spectra caused by the large rotational velocities of the components. Therefore, we decided to fix the luminosity ratio to 1 based on the results found in Sects.\,\ref{sec:V402Lac} and \ref{sec:V402LacLC}, which point towards two equal-mass stars of similar luminosities. $T_{\rm eff}$ and $v_{\rm b}$ values derived for V402\,Lac, with the adopted $L_2/L_1=1$, are listed in Table\,\ref{tab:todmorV402}.

Finally, RV curves were computed consistently with \texttt{todmor} using as templates for each instrument the ones with their parameters closer to those listed in Tables~\ref{tab:todmorV889} and \ref{tab:todmorV402}. The RVs of spectra taken during any of the eclipses or at orbital phases close to  conjunction, where the spectral lines of the components could not be separated due to rotational broadening, were not further considered. The final number of RVs used in this analysis is shown in parenthesis in Table~\ref{tab:specdata}. We list the derived RVs of each object in Tables~\ref{tab:V889AqlRVs} and \ref{tab:V402LacRVs}. 

Concerning the possibility to check rotational velocities using the Rossiter-McLaughlin effect \citep{Rossiter1924,McLaughlin1924}, we could obtain eight CAFE spectra within the primary eclipse of V889\,Aql but not for the shorter period case of V402\,Lac. Observations showed that the rotational effect on the radial velocities during eclipse could not provide any indication of misalignment of the rotational axes of V889\,Aql, although more observations during both eclipses are needed to confirm the actual alignment of all components.

\begin{table}[!t]
\centering
\caption{Template parameters of V889\,Aql found from the maximisation of the CCF peak of all available spectra with \texttt{todmor} and their weighted means with uncertainties.}
\label{tab:todmorV889}
\begin{tabular}{l cccc} 
\hline\hline
\noalign{\smallskip}
\multirow{2}{*}{Instrument} & $T_{\rm eff,1}$ & $T_{\rm eff,2}$ & $v_{\rm b,1}$ & $v_{\rm b,2}$  \\
\noalign{\smallskip}
 & [K] & [K] & [km\,s$^{-1}$] & [km\,s$^{-1}$] \\
\noalign{\smallskip}
\hline
\noalign{\smallskip}
CAFE & $9240\pm100$ & $9200\pm90$ & $15.8\pm0.5$ & $20.1\pm0.8$ \\
TIGRE & $9040\pm100$ & $8920\pm150$ & $19.9\pm0.4$ & $23.4\pm1.0$  \\
Mean & $9140\pm100$ & $9130\pm110$ & $18.1\pm0.5$ & $21.4\pm0.9$ \\
\noalign{\smallskip}
\hline
\end{tabular}
\end{table}

\begin{table}[!t]
\centering
\caption{Same as Table \ref{tab:todmorV889} for V402\,Lac.}
\label{tab:todmorV402}
\begin{tabular}{l cccc} 
\hline\hline
\noalign{\smallskip}
\multirow{2}{*}{Instrument} & $T_{\rm eff,1}$ & $T_{\rm eff,2}$ & $v_{\rm b,1}$ & $v_{\rm b,2}$  \\
\noalign{\smallskip}
 & [K] & [K] & [km\,s$^{-1}$] & [km\,s$^{-1}$] \\
\noalign{\smallskip}
\hline
\noalign{\smallskip}
CAFE & $11400\pm200$ & $11200\pm200$ & $108\pm6$ & $111\pm8$ \\
Tull & $11200\pm300$ & $11100\pm200$ & $109\pm5$ & $108\pm5$  \\
STELLA & $11600\pm400$ & $11200\pm200$ & $112\pm12$ & $111\pm12$  \\
Mean & $11400\pm300$ & $11200\pm200$ & $109\pm7$ & $109\pm7$ \\
\noalign{\smallskip}
\hline
\end{tabular}
\end{table}

\subsection{Photometry}

The two systems studied in this work have been observed by the Transiting Exoplanet Survey Satellite ({\em TESS}) mission \citep{Ricker2015}, which is an all-sky photometric survey obtaining densely-covered light curves at very high precision of the nearest and brightest stars, aimed at detecting exoplanets. The excellent quality of {\em TESS} data make them ideal for the study of stellar eclipse events \citep{Baroch2021,Claret2021,Southworth2021b}, which are much more prominent than exoplanet transits.

\begin{figure}[!t]
    \centering
    \includegraphics[width=\columnwidth]{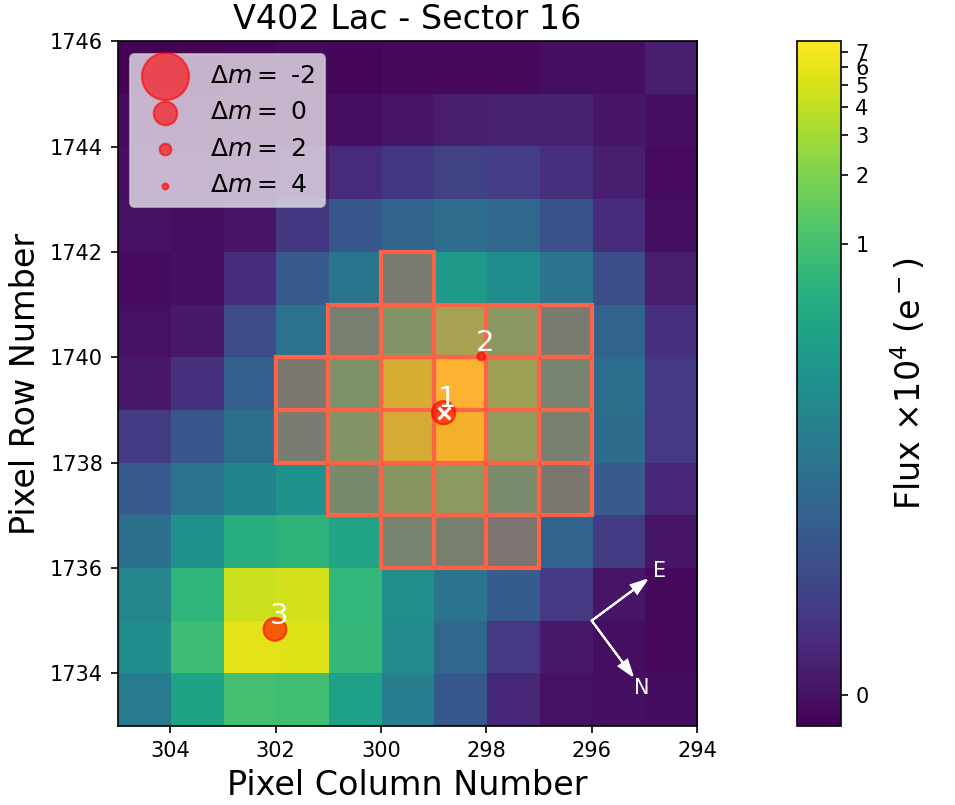}
    \caption{{\em TESS} target pixel file of V402\,Lac in Sector 16 \citep[created with \texttt{tpfplotter},][]{Aller2020}, with the colour code indicating the flux measured by each pixel. We show the pixels included in the computation of the SAP as orange bordered pixels. The size of the red circles indicates the {\em TESS} magnitudes of the three brightest stars contained in the field, with V402\,Lac marked with a white cross.}
    \label{fig:V402mask}
\end{figure}

V402\,Lac was one of the {\em TESS} targets included in Sector 16, which was observed during September 2019. We collected the 2-minute cadence simple aperture photometry (SAP) produced by the Science Process Operation Centre \citep[SPOC,][]{Jenkins2016}, available at the Mikulski Archive for Space Telescopes\footnote{https://mast.stsci.edu/portal/Mashup/Clients/Mast/Portal.html} (MAST). We show in Fig.\,\ref{fig:V402mask} the mask used for the computation of the SAP and the brightest neighbours to V402\,Lac. As it can be seen in the Figure, the mask contains a star (BD+44 4060, label \#2) with {\em Gaia} magnitude $G=9.87$, which we estimate to contribute 5.2\% of the total flux in the {\em Gaia} band. In addition, the bright visual companion to V402\,Lac, HD\,210387  (label \#3, $G=6.72$), has a PSF extending several pixels inside the mask, which would also be significantly contributing to the total flux through an undetermined amount.

\begin{figure}[!t]
    \centering
    \includegraphics[width=\columnwidth]{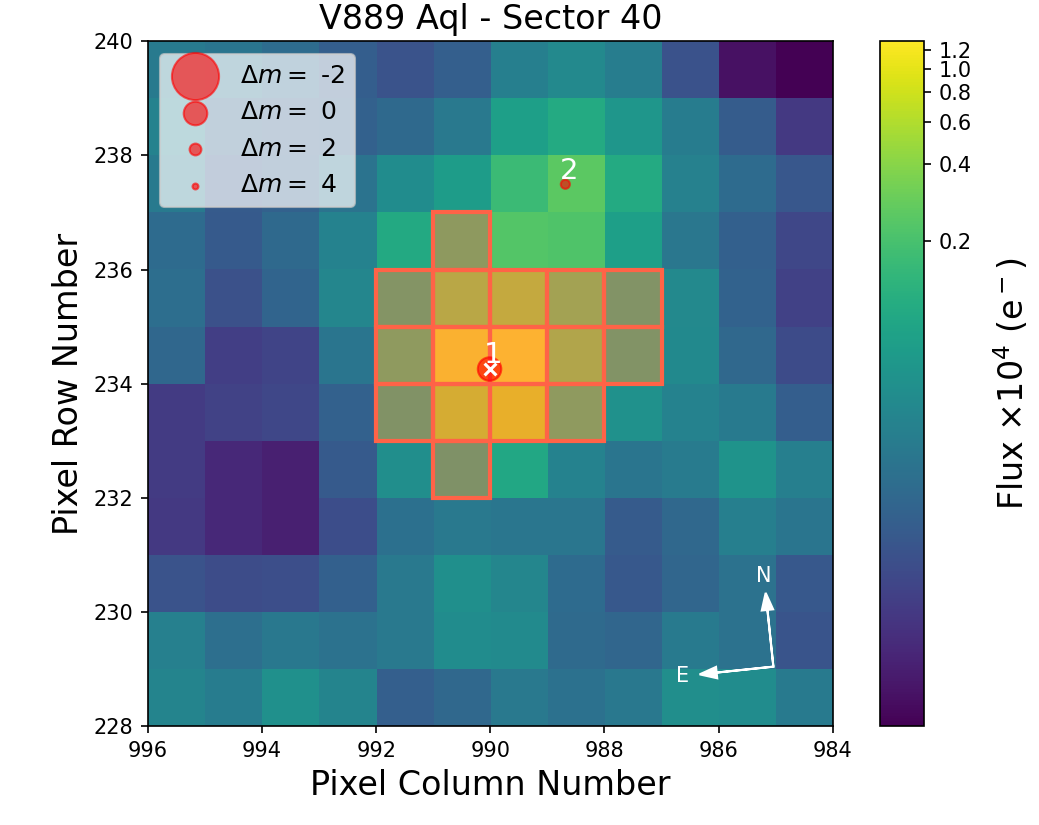}
    \caption{Portion of the {\em TESS} FFI containing V889\,Aql in Sector 40 \citep[created with \texttt{tpfplotter},][]{Aller2020}, with the colour code indicating the flux measured by each pixel. We show the pixels used to extract the photometry as orange bordered pixels. The size of the red circles indicates the {\em TESS} magnitudes of the two brightest stars contained in the field, with V889\,Aql marked with a white cross.}
    \label{fig:V889mask}
\end{figure}

On the other hand, V889\,Aql was not included as a {\em TESS} target in any of the sectors, though it fell within the {\em TESS} field of view during Sectors 14 and 40, which were observed between July and August 2019 and between June and July 2021, with a cadence of 30 and 10 minutes, respectively. Therefore, we downloaded the {\em TESS} full frame images (FFIs) from MAST, and used the public {\em TESS} aperture photometry tool \texttt{Eleanor}\footnote{https://adina.feinste.in/eleanor/} \citep{Feinstein2019} to extract the SAP of V889\,Aql with a mask defined by the pixels shown in Fig.\,\ref{fig:V889mask}. The mask was modified to avoid the addition of a visual companion (label \#2, $G=11.21$), which was included in the default mask computed by \texttt{Eleanor}.

\begin{figure}[!t]
    \centering
    \includegraphics[width=\columnwidth]{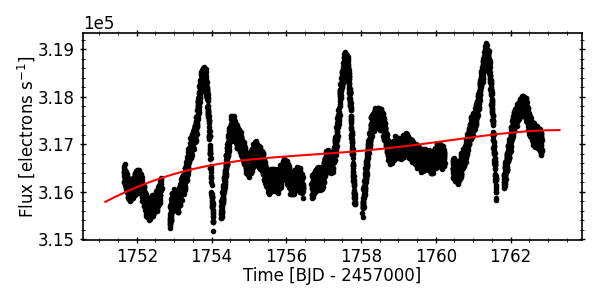}
    \caption{Model (red) used to detrend the out-of-eclipse {\em TESS} photometry (black dots) of V402\,Lac in the second part of Sector 16.}
    \label{fig:V402TESS}
\end{figure}

\begin{figure}[!t]
    \centering
    \includegraphics[width=\columnwidth]{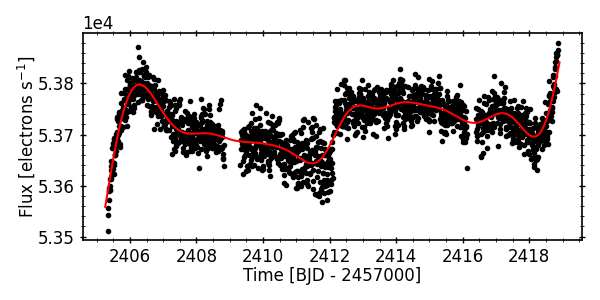}
    \caption{Model (red) used to detrend the out-of-eclipse {\em TESS} photometry (black dots) of V889\,Aql in the second part of Sector 40.}
    \label{fig:V889TESS}
\end{figure}

The out-of-eclipse {\em TESS} photometry of both stars, shown in  Figs.\,\ref{fig:V402TESS} and \ref{fig:V889TESS}, present long term trends produced by effects such as, e.g., pointing jitter and scattered light from the Earth and Moon \citep{Hattori2021}, which can bias the resulting orbital parameters and measured times of minimum light. To mitigate the effects of these trends, we employed the Python package \texttt{george} \citep{george} to model the out-of-eclipse photometry using a Gaussian process correlated-noise model with a squared-exponential covariance function \citep[see e.g.][for more details]{Gibson2012,Aigrain2016}. The length-scale hyperparameter was set to a value of 2 days for V889\,Aql, while for V402\,Lac was set to a value of 20 days to avoid fitting the modulations produced by geometric effects, observed with a periodicity of 3.78\,d, the orbital period of the system. The resulting model, shown as red lines in the upper panels in Figs.\,\ref{fig:V402TESS} and \ref{fig:V889TESS}, were used to normalise the entire light curve, including the eclipses. 

In the photometric analysis presented in Sect.\,\ref{sec:dimensions}, in order to speed up the optimisation of the orbital parameters, we reduced the number of photometric points by only using the last part (i.e., after intermediate data downlink) of the {\em TESS} lightcurve in Sector 16 for V889\,Aql and Sector 40 in V402\,Lac, which in both cases seem to be the portions less affected by {\em TESS} systematics. The data used in the analysis represent $\sim$37\% of the available data for V889\,Aql, containing one pair of primary and secondary eclipses, and a $\sim$50\,\% of the available data for V402\,Lac, containing 3 pairs of primary and secondary eclipses.

\subsection{Times of minimum light}

Following the same approach as in \cite{Baroch2021}, we determined the time of minimum light of the {\em TESS} photometry using the \cite{Kwee1956} method (KvW), considering only eclipses where both the ingress and egress are well sampled, and using the same orbital phase interval for all primary and secondary eclipses for consistency. We discarded eclipses with strong asymmetries in their bisectors \citep[see][for more details]{Baroch2021}, which could be indicative of biased timing determinations due to, e.g., systematic trends remaining in the data after our modelling. As a result of this procedure, we obtained 7 useful eclipse timings of V889\,Aql (4 primary and 3 secondary) and 12 of V402\,Lac (6 primary and 6 secondary).

In addition, we collected numerous photoelectric times of eclipse of V889\,Aql and V402\,Lac that have been measured over the last decades, and listed them in Tables\,\ref{tab:V889times} and \ref{tab:V402times}. In total, 40 eclipse timings (20 primary and 20 secondary) are available for V889\,Aql and 50 (31 primary and 19 secondary) for V402\,Lac.

Preliminary attempts to fit the ephemeris of the eclipse timings of V402\,Lac yielded residuals with a median deviation ten times larger than the published uncertainties, with some timings deviating up to 100 times the stated error. Although several works have shown that in the presence of systematic deviations from an ideal eclipse the KvW method may produce exceedingly optimistic errors \citep{Breinhorst1973,Mikulasek2014,Torres2017}, the deviations are much larger that those usually observed (cf. the timings of V889\,Aql have a median deviation $\sim$1.5 times larger than their errors). For this reason, we decided to use errors of 0.005 and 0.010 days to primary and secondary timings, respectively, and errors of 0.0001 and 0.0004 days to primary and secondary {\em TESS} timings, respectively. These values are set to approximately match the dispersion shown by the residuals of the fit described in Sect.~\ref{sec:V402Lac}.

\begin{table}[!t]
\centering
\caption{Times of eclipse for V889\,Aql.}
\label{tab:V889times}
\begin{tabular}{llcl} 
\hline\hline
\noalign{\smallskip}
BJD & Error & \multirow{2}{*}{Type} & \multirow{2}{*}{Reference} \\
$(+2400000)$ & [d] & & \\
\noalign{\smallskip}
\hline
\noalign{\smallskip}
38164.4901 & 0.0005 & 1 & \cite{Semeniuk1967} \\
38242.3352 & 0.0006 & 1 & \cite{Semeniuk1967} \\
38257.4204 & 0.0010 & 2 & \cite{Semeniuk1967} \\
40088.396 & 0.005\tablefootmark{a} & 1 & \cite{Bernardi1978} \\
41927.306 & 0.005\tablefootmark{a} & 2 & \cite{Bernardi1978} \\
42568.359 & 0.005\tablefootmark{a} & 1 & \cite{Bernardi1978} \\
42583.429 & 0.005\tablefootmark{a} & 2 & \cite{Bernardi1978} \\
42594.550 & 0.005\tablefootmark{a} & 2 & \cite{Bernardi1978} \\
42961.549 & 0.005\tablefootmark{a} & 2 & \cite{Bernardi1978} \\
43246.731 & 0.005\tablefootmark{a} & 1 & \cite{Bernardi1978} \\
43250.694 & 0.005\tablefootmark{a} & 2 & \cite{Bernardi1978} \\
44047.433 & 0.005\tablefootmark{a} & 1 & \cite{Gimenez1982} \\
44062.507 & 0.005\tablefootmark{a} & 2 & \cite{Gimenez1982} \\
44340.535 & 0.005\tablefootmark{a} & 2 & \cite{Gimenez1982} \\
44492.270 & 0.005\tablefootmark{a} & 1 & \cite{Gimenez1982} \\
44785.364 & 0.005\tablefootmark{a} & 2 & \cite{Khaliullin1989} \\
44803.656 & 0.005\tablefootmark{a} & 1 & \cite{Skillman1982} \\
45515.391 & 0.005\tablefootmark{a} & 1 & \cite{Khaliullin1989} \\
46264.442 & 0.005\tablefootmark{a} & 2 & \cite{Diethelm1985} \\
48106.537 & 0.005\tablefootmark{a} & 1 & \cite{Diethelm1990} \\
48755.483 & 0.002 & 2 & \cite{Hegedus1996} \\
50312.388 & 0.005\tablefootmark{a}  & 2 & \cite{Diethelm1996} \\
53255.3928 & 0.0010 & 1 & \cite{Biro2007} \\
53993.2958 & 0.0005 & 2 & \cite{Bozkurt2011} \\
54645.4737 & 0.0006 & 1 & \cite{Brat2008} \\
54938.5513 & 0.0010 & 2 & \cite{Brat2009} \\
55012.4554 & 0.0006 & 1 & \cite{Brat2011}\\
55835.3860 & 0.0004 & 1 & \cite{Honkova2013} \\
56072.8595 & 0.0008 & 2 & \cite{Diethelm2012} \\
56491.5066 & 0.0003 & 1 & \cite{Diethelm2014} \\
56495.442 & 0.005\tablefootmark{a} & 2 & \cite{Braune2014}\\
57240.536 & 0.006 & 2 & \cite{Hubscher2017} \\
57592.456 & 0.004 & 1 & \cite{Hubscher2017} \\
58686.22397 & 0.00016 & 2 & This work \\
58693.41033 & 0.00007 & 1 & This work \\
58704.5310 & 0.0002 & 1 & This work \\
58708.46522 & 0.00011 & 2 & This work \\
59394.01831 & 0.00005 & 1 & This work \\
59409.07196 & 0.00005 & 2 & This work \\
59416.25984 & 0.00002 & 1 & This work \\
\noalign{\smallskip}
\hline
\end{tabular}
\tablefoot{Primary and secondary eclipses are labelled as 1 and 2 in the Type column, respectively.
\tablefoottext{a}{Timing with no published error, adopted to be 0.005 days.}
}
\end{table}

\begin{table}[!t]
\centering
\caption{Times of eclipse for V402\,Lac.}
\label{tab:V402times}
\begin{tabular}{llcl} 
\hline\hline
\noalign{\smallskip}
BJD & Error & \multirow{2}{*}{Type} & \multirow{2}{*}{Reference} \\
$(+2400000)$ & [d] & & \\
\noalign{\smallskip}
\hline
\noalign{\smallskip}
50955.501  & 0.005 & 1 & \cite{Herrero2010} \\
51366.561  & 0.010 & 2 & \cite{Herrero2010} \\
51371.520  & 0.005 & 1 & \cite{Herrero2010} \\
51839.304  & 0.010 & 2 & \cite{Hegedus2003} \\
52115.405  & 0.010 & 2 & \cite{Bulut2003}  \\
52116.573  & 0.005 & 1 & \cite{Bulut2003}  \\
53900.505  & 0.010 & 2 & \cite{Herrero2010} \\
53905.533  & 0.005 & 1 & \cite{Herrero2010} \\
55051.503  & 0.005 & 1 & \cite{Hubscher2010} \\
55081.760  & 0.005 & 1 & \cite{Herrero2010} \\
55085.544  & 0.005 & 1 & \cite{Herrero2010} \\
55118.310  & 0.010 & 2 & \cite{Herrero2010} \\
55123.363  & 0.005 & 1 & \cite{Herrero2010} \\
55380.544 &  0.005 & 1 & \cite{Agusti2012} \\
55433.494 &  0.005 & 1 & \cite{Brat2011} \\
55443.571 &  0.010 & 2 & \cite{Agusti2012} \\
55443.562 &  0.010 & 2 & \cite{Brat2011} \\
55448.622 &  0.005 & 1 & \cite{Agusti2012} \\
55496.503 &  0.010 & 2 & \cite{Agusti2012} \\
55500.289 &  0.010 & 2 & \cite{Zasche2011} \\
55500.295 &  0.010 & 2 & \cite{Zasche2011}  \\
55505.354 &  0.010 & 1 & \cite{Agusti2012} \\
55806.612 &  0.010 & 2 & \cite{Agusti2012} \\
55815.482 &  0.005 & 1 & \cite{Zasche2011}  \\
55825.534 &  0.010 & 2 & \cite{Agusti2012} \\
56163.434 &  0.005 & 1 & \cite{Zasche2014} \\
56511.383 &  0.005 & 1 & \cite{Zasche2014} \\
56564.335 &  0.005 & 1 & \cite{Hoyman2018} \\
56821.516 &  0.005 & 1 & \cite{Zasche2014} \\
56874.462 &  0.005 & 1 & \cite{Hoyman2018} \\
56893.374 &  0.005 & 1 & \cite{Hoyman2018} \\
57179.482 &  0.010 & 2 & \cite{Zasche2017} \\
57203.506 &  0.005 & 1 & \cite{Zasche2017} \\
57237.544 &  0.005 & 1 & \cite{Hubscher2017} \\
58330.562 &  0.005 & 1 & \cite{Pagel2020} \\
58402.422 &  0.005 & 1 & \cite{Paschke2019} \\
58731.463 &  0.005 & 1 & \cite{Pagel2020}\\
58739.02667 &  0.00010 & 1 & This work \\
58741.4330 &  0.0004 & 2 & This work \\
58742.80886 &  0.00010 & 1 & This work \\
58745.2140 &  0.0004 & 2 & This work \\
58746.59120 &  0.00010 & 1 & This work \\
58748.9973 &  0.0004 & 2 & This work \\
58752.7785 &  0.0004 & 2 & This work \\
58754.15512 &  0.00010 & 1 & This work \\
58756.5605 &  0.0004 & 2 & This work \\
58757.93735 &  0.00010 & 1 & This work \\
58760.3428 &  0.0004 & 2 & This work \\
58761.71929 &  0.00010 & 1 & This work \\
\noalign{\smallskip}
\hline
\end{tabular}
\tablefoot{Primary and secondary eclipses are labelled as 1 and 2 in the Type column, respectively. We assigned uncertainties of 0.005 and 0.010 days to primary and secondary literature timings, and 0.0001 and 0.0004 days to primary and secondary {\em TESS} timings, see text. 
}
\end{table}

\section{Study of the orbital parameters}
\label{sec:orbital}

For the determination of the orbital parameters from RVs we usually need good knowledge of the orbital period, which is generally constrained to high precision from the times of minimum light. At the same time, the analysis of the eclipse timings in eccentric orbits require good determination of the orbital eccentricity, which can be highly correlated with the apsidal motion \citep{Torres2015}. Therefore, a combined solution of the radial velocities and the times of minimum was obtained, allowing a precise simultaneous determination of the anomalistic orbital period and the eccentricity. This procedure has been already successfully followed in the somewhat similar binary systems V501\,Mon \citep{Torres2015} and V541\,Cyg \citep{Torres2017}.

For the RV model we assumed a Keplerian orbital motion, and solved for the sidereal period ($P_{\rm s}$), the semi-amplitudes of the components ($K_{\rm 1}$ and $K_{\rm 2}$), the eccentricity ($e$), the apsidal motion ($\dot{\omega}$), the argument of periastron ($\omega_0$) and time of periastron passage ($T_{\rm peri}$) defined at a reference time of the primary eclipse $T_{\rm I}$, and the systemic velocity of each instrument ($\gamma_{\rm Inst.}$). Additionally, we also allowed an adjustable RV jitter term for each instrument (Jit$_{\rm Inst.}$), which was added in quadrature to the errors as described in \cite{Baluev2009}. Simultaneously, the times of eclipse were modelled using Eq.~11 in \cite{Gimenez1995}, solving for $P_{\rm a}$, $T_{\rm I}$, $e$, $\omega_0$ and $\dot{\omega}$. We also modelled the light travel time effect over the times of eclipse due to a potential third body with Eq.~2 of \cite{Irwin1959}. This additional effect yields the orbital parameters of the third body, as well as the time semi-amplitude ($\tau$), from which we can estimate the minimum mass.

The optimisation of all orbital parameters and computation of uncertainties was done by sampling the posterior probability distribution with the \texttt{emcee} sampler \citep{Foreman2013}, an implementation of the affine-invariant ensemble sampler Markov chain Monte Carlo \citep{Goodman2010}.

\begin{table}[!t]
\centering
\caption{Parameters of the best joint fit to the radial velocities and eclipse timings of V889\,Aql.}
\label{tab:V889results}
\begin{tabular}{l c} 
\hline\hline
\noalign{\smallskip}
 Parameter & Value  \\
\noalign{\smallskip}
\hline
\noalign{\smallskip}
\multicolumn{2}{c}{\textit{Inner binary}} \\
\noalign{\smallskip}
\hline
\noalign{\smallskip}
$P_{\rm a}$ [d] & $11.120771\pm0.000004$ \\
\noalign{\smallskip}
$P_{\rm s}$ [d] & $11.120757\pm0.000004$ \\
\noalign{\smallskip}
$T_{\rm I}$ & $2459416.2598\pm0.00002$ \\
\noalign{\smallskip}
$T_{\rm II}$ & $2459420.19276\pm0.00006$ \\
\noalign{\smallskip}
$T_{\rm peri}$ & $2459416.752\pm0.006$ \\
\noalign{\smallskip}
$e$ & $0.3750\pm0.0013$ \\
\noalign{\smallskip}
$\omega_0$ [deg] & $126.1\pm0.2$ \\
\noalign{\smallskip}
$e\sin \omega_0$ & $0.303\pm0.002$ \\
\noalign{\smallskip}
$e\cos \omega_0$ & $-0.22107\pm0.00013$ \\
\noalign{\smallskip}
$\dot{\omega}$ [deg\,cycle$^{-1}$] & $0.00046\pm0.00002$ \\
\noalign{\smallskip}
$K_{\rm A}$ [km\,s$^{-1}$] & $82.7\pm0.3$ \\
\noalign{\smallskip}
$K_{\rm B}$ [km\,s$^{-1}$] & $84.6\pm0.3$ \\
\noalign{\smallskip}
$M_{\rm A} \sin^3 i$ [M$_{\odot}$] & $2.17\pm0.02$ \\
\noalign{\smallskip}
$M_{\rm B} \sin^3 i$ [M$_{\odot}$] & $2.125\pm0.014$ \\
\noalign{\smallskip}
$q$ & $0.978\pm0.005$ \\
\noalign{\smallskip}
$a\sin i$ [au] & $0.1585\pm0.0003$ \\
\noalign{\smallskip}
$\gamma_{\rm CAFE}$ [km\,s$^{-1}$]& $-20.9\pm0.2$ \\
\noalign{\smallskip}
$\gamma_{\rm TIGRE}$ [km\,s$^{-1}$] & $-19.1\pm0.2$ \\
\noalign{\smallskip}
Jit$_{\rm CAFE}$ [km\,s$^{-1}$] & $0.2\pm0.2$ \\
\noalign{\smallskip}
Jit$_{\rm TIGRE}$ [km\,s$^{-1}$] & $0.3\pm0.3$ \\
\noalign{\smallskip}
\hline
\noalign{\smallskip}
\multicolumn{2}{c}{\textit{Third body}} \\
\noalign{\smallskip}
\hline
\noalign{\smallskip}
$P_{\rm 3}$ [a] & $67\pm4$ \\
\noalign{\smallskip}
$T_{\rm peri,3}$ & $2447700\pm1200$ \\
\noalign{\smallskip}
$\tau_{\rm 3}$ [d] & $0.047\pm0.002$ \\
\noalign{\smallskip}
$e_{\rm 3}$ & $0.24\pm0.04$ \\
\noalign{\smallskip}
$\omega_{\rm 3}$ [deg] & $100\pm20$ \\
\noalign{\smallskip}
$f(m_{\rm 3})$ [M$_{\odot}$] & $0.122\pm0.012$ \\
\noalign{\smallskip}
$a_{\rm 3} \sin i$ [au] & $8.2\pm0.4$ \\
\noalign{\smallskip}
\hline
\end{tabular}
\end{table}

\subsection{V889\,Aql} \label{sec:V889Aql}

\begin{figure}[!t]
    \centering
    \includegraphics[width=\columnwidth]{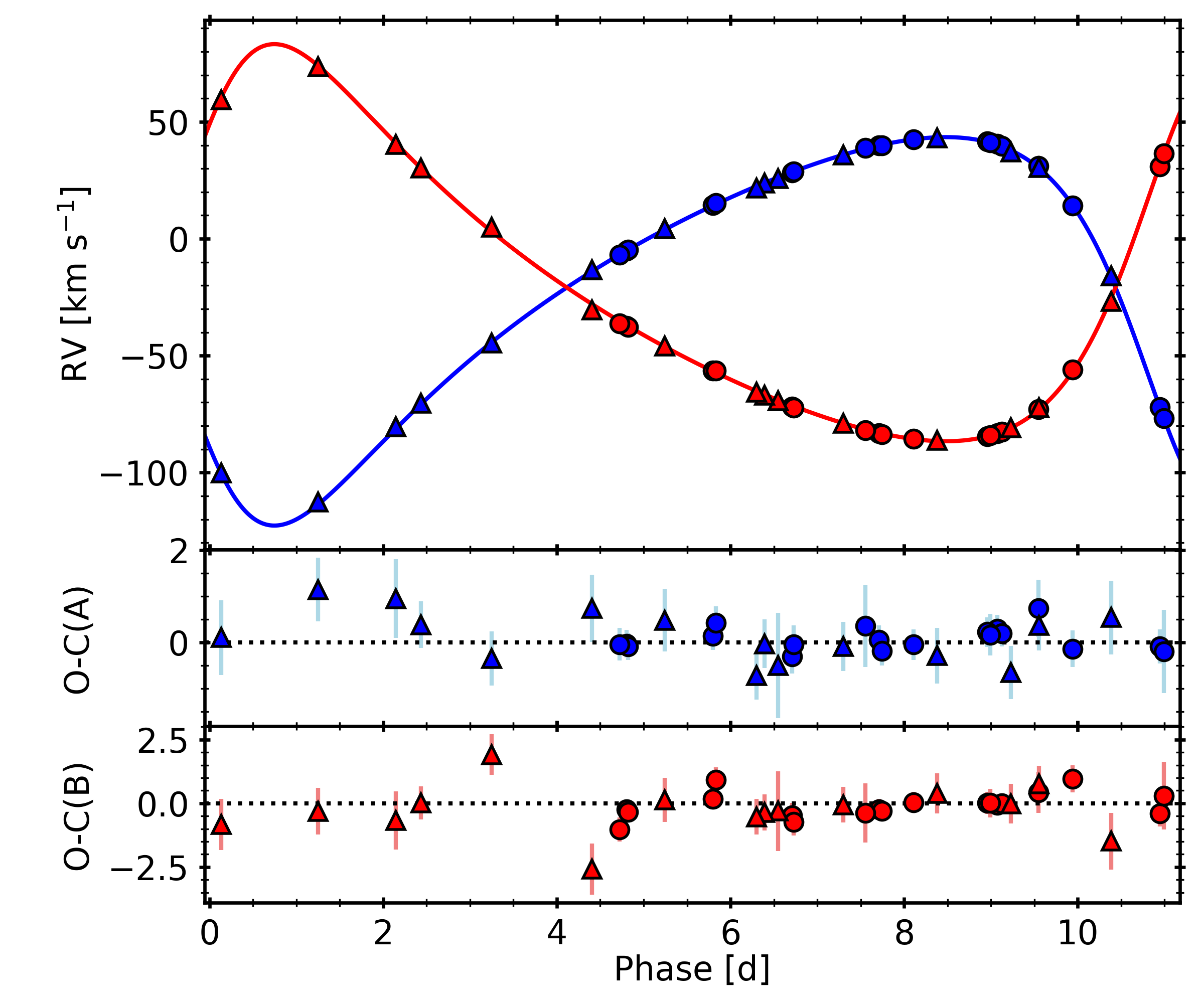}
    \caption{Best fit to the CAFE (circles) and TIGRE (triangles) RVs of V889\,Aql, phase-folded to the orbital period. Blue and red colours correspond to the primary and secondary components, respectively. The bottom panels show the residuals of the best fit.}
    \label{fig:V889Aql_RV}
\end{figure}

\begin{figure*}[!t]
    \centering
    \includegraphics[width=0.9\textwidth]{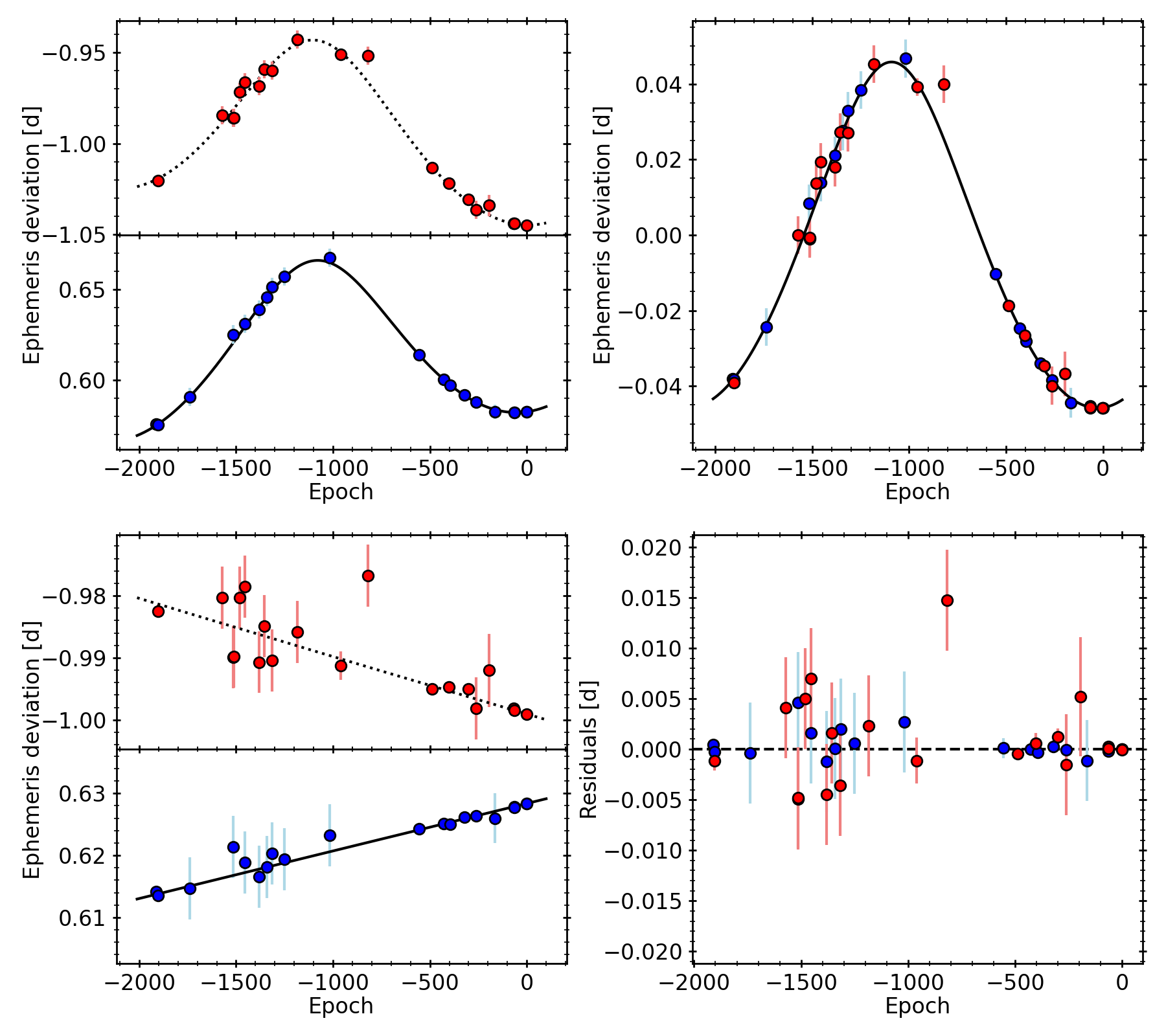}
    \caption{Best fit to the primary (blue) and secondary (red) times of minimum light of V889\,Aql, after subtracting linear ephemeris (top-left), after subtracting linear ephemeris and the effect of the third body (bottom-left), and after subtracting linear ephemeris and the apsidal motion (top-right). The panel at the bottom-right shows the residuals from all the fits.}
    \label{fig:V889Aql_OC}
\end{figure*}

We used the model described above to simultaneously fit the RVs in Table\,\ref{tab:V889AqlRVs} and eclipse timings in Table\,\ref{tab:V889times}. We show the best fits in Figs.\,\ref{fig:V889Aql_RV} and \ref{fig:V889Aql_OC}. The parameters of the best joint fit are given in Table\,\ref{tab:V889results}. The value of the orbital inclination, needed to fit the eclipse timings with eq.~11 in \cite{Gimenez1995}, was adopted from the preliminary solution of the light curve to be 89\,deg (see section 4.1), and the fitted parameters were found to be very robust to variations of the adopted value. As expected, the simultaneous fit of both eclipse timings and RVs allowed us to obtain tight constrains on the anomalistic orbital period and the eccentricity, with $P_{\rm a}=11.120771\pm0.000004$\,d and $e=0.3750\pm0.0013$, which otherwise would be impossible to determine without fixing one of the values. The velocity semi-amplitude values obtained in the best fit indicate that the two stars have very similar masses, but not identical, with a mass ratio $q = 0.978 \pm 0.005$. In addition, we also obtained a precise determination of the apsidal motion, of $\dot{\omega}=0.00046\pm0.00002$\,deg\,cycle$^{-1}$, in agreement with previous determinations \citep{Gimenez1982,Khaliullin1989,Wolf2005} and consistent within mutual uncertainties with the value of $0.000487\pm0.000013$\,deg\,cycle$^{-1}$ obtained by \citep{Kiran2019}, who modelled the eclipse timings with a free value for the eccentricity. The effect of the apsidal motion on the eclipse timings is clearly seen in Fig.\,\ref{fig:V889Aql_OC}. In the upper-left panel, we show the eclipse timings once the linear ephemeris \cite[i.e. the first three terms in eq.\,11 in][]{Gimenez1995} are subtracted, leaving only the effects of the apsidal motion and the third body, while the lower-left panel show the effect of the apsidal motion alone (i.e. after the linear ephemeris and the third body effect are subtracted).

Similarly, in the upper-right panel in Fig.\,\ref{fig:V889Aql_OC} we show the eclipse timings after linear ephemeris and apsidal motion contribution are subtracted, leaving a clear indication of the presence of a third body around V889\,Aql, as already found by \cite{Wolf2005} and \cite{Kiran2019}. The best-fit parameters indicate a third body orbital period of $67\pm4$\,a and an eccentricity of $0.24\pm0.04$. The third body induces a semi-amplitude over the eclipse timing residuals from a linear ephemeris of $0.047\pm0.002$\,d, from which we derive a mass function of $0.122\pm0.012$\,M$_{\odot}$, or equivalently, a minimum mass of $\sim1.6$\,M$_{\odot}$.

As a sanity check to confirm that the presence of a third body is not affecting the measured value of the apsidal motion term, we used eq.\,7 in \cite{Baroch2021} to compute the apsidal motion from a linear fit to the difference between secondary and primary eclipse timings ($T_2-T_1$), from which we derive an apsidal motion of $\dot{\omega}=0.00047\pm0.00002$\,deg\,cycle$^{-1}$, confirming the value found previously.

\subsection{V402\,Lac} \label{sec:V402Lac}

Following the same approach as for V889\,Aql, we simultaneously modelled the RVs and eclipse timings of V402\,Lac, listed in Tables\,\ref{tab:V402LacRVs} and \ref{tab:V402times}, respectively. In this case, however, the RV observations span almost 20 years, covering a non-negligible portion of the apsidal motion period, which we expect to cause a significant deviation of the RVs from a Keplerian orbit due to periastron precession. For this reason, we modelled the RVs with a Keplerian orbit, but allowing the argument of periastron, $\omega$, to vary according to the apsidal motion rate $\dot{\omega}$ as $\omega=\omega_0+\dot{\omega}\cdot N$, with $N$ representing the number of cycles since $T_{\rm I}$ and with $\omega_0$ defined at the same reference time of primary eclipse $T_{\rm I}$ used in the fit to the eclipse timings. We list the obtained best fitting parameters of the joint model in Table\,\ref{tab:V402results}. We obtained an anomalistic orbital period $P_{\rm a}=3.782138\pm0.000006$\,d with an eccentricity of $e=0.376\pm0.003$. The obtained values of the semi-amplitudes are compatible within the uncertainties, indicating that the two component stars are nearly indistinguishable, with a mass ratio of $q=0.995\pm0.008$. The improvement in the precision of the eccentricity value with respect to other works \citep{Bulut2013,Hoyman2018} allows us to precisely determine the apsidal motion rate of the system, yielding a value of $\dot{\omega}=0.0090\pm0.0003$\,deg\,cycle$^{-1}$. 

\begin{table}[!t]
\centering
\caption{Parameters of the best joint fit to the radial velocities and eclipse timings of V402\,Lac.}
\label{tab:V402results}
\begin{tabular}{l c} 
\hline\hline
\noalign{\smallskip}
 Parameter & Value  \\
\noalign{\smallskip}
\hline
\noalign{\smallskip}
$P_{\rm a}$ [d] & $3.782138\pm0.000006$ \\
\noalign{\smallskip}
$P_{\rm s}$ [d] & $3.782043\pm0.000005$ \\
\noalign{\smallskip}
$T_{\rm I}$ & $2458761.71931\pm0.00006$ \\
\noalign{\smallskip}
$T_{\rm II}$ & $2458764.1245\pm0.0002$ \\
\noalign{\smallskip}
$T_{\rm peri}$ & $2458761.579\pm0.008$ \\
\noalign{\smallskip}
$e$ & $0.376\pm0.003$ \\
\noalign{\smallskip}
$\omega$ [deg] & $57.9\pm0.4$ \\
\noalign{\smallskip}
$e\sin \omega$ & $0.319\pm0.004$ \\
\noalign{\smallskip}
$e\cos \omega$ & $0.1996\pm0.0003$ \\
\noalign{\smallskip}
$\dot{\omega}$ [deg\,cycle$^{-1}$] & $0.0090\pm0.0003$ \\
\noalign{\smallskip}
$K_{\rm A}$ [km\,s$^{-1}$] & $128.5\pm0.8$ \\
\noalign{\smallskip}
$K_{\rm B}$ [km\,s$^{-1}$] & $129.2\pm0.8$ \\
\noalign{\smallskip}
$M_{\rm A} \sin^3 i$ [M$_{\odot}$] & $2.67\pm0.04$ \\
\noalign{\smallskip}
$M_{\rm B} \sin^3 i$ [M$_{\odot}$] & $2.66\pm0.04$ \\
\noalign{\smallskip}
$q$ & $0.995\pm0.008$ \\
\noalign{\smallskip}
$a\sin i$ [au] & $0.0830\pm0.0003$ \\
\noalign{\smallskip}
$\gamma_{\rm CAFE}$ [km\,s$^{-1}$]& $-4.8\pm1.0$ \\
\noalign{\smallskip}
$\gamma_{\rm Tull}$ [km\,s$^{-1}$] & $1.9\pm0.7$ \\
\noalign{\smallskip}
$\gamma_{\rm STELLA}$ [km\,s$^{-1}$] & $-18.3\pm0.7$ \\
\noalign{\smallskip}
Jit$_{\rm CAFE}$ [km\,s$^{-1}$] & $4.0\pm1.0$ \\
\noalign{\smallskip}
Jit$_{\rm Tull}$ [km\,s$^{-1}$] & $3.1\pm0.7$ \\
\noalign{\smallskip}
Jit$_{\rm STELLA}$ [km\,s$^{-1}$] & $3.2\pm0.7$ \\
\noalign{\smallskip}
\hline
\noalign{\smallskip}
\multicolumn{2}{c}{\textit{Third body}} \\
\noalign{\smallskip}
\hline
\noalign{\smallskip}
$P_{\rm 3}$ [a] & $>20$ \\
\noalign{\smallskip}
$\tau_{\rm 3}$ [d] & $>0.004$ \\
\noalign{\smallskip}
\hline
\end{tabular}
\end{table}

\begin{figure}
    \centering
    \includegraphics[width=\columnwidth]{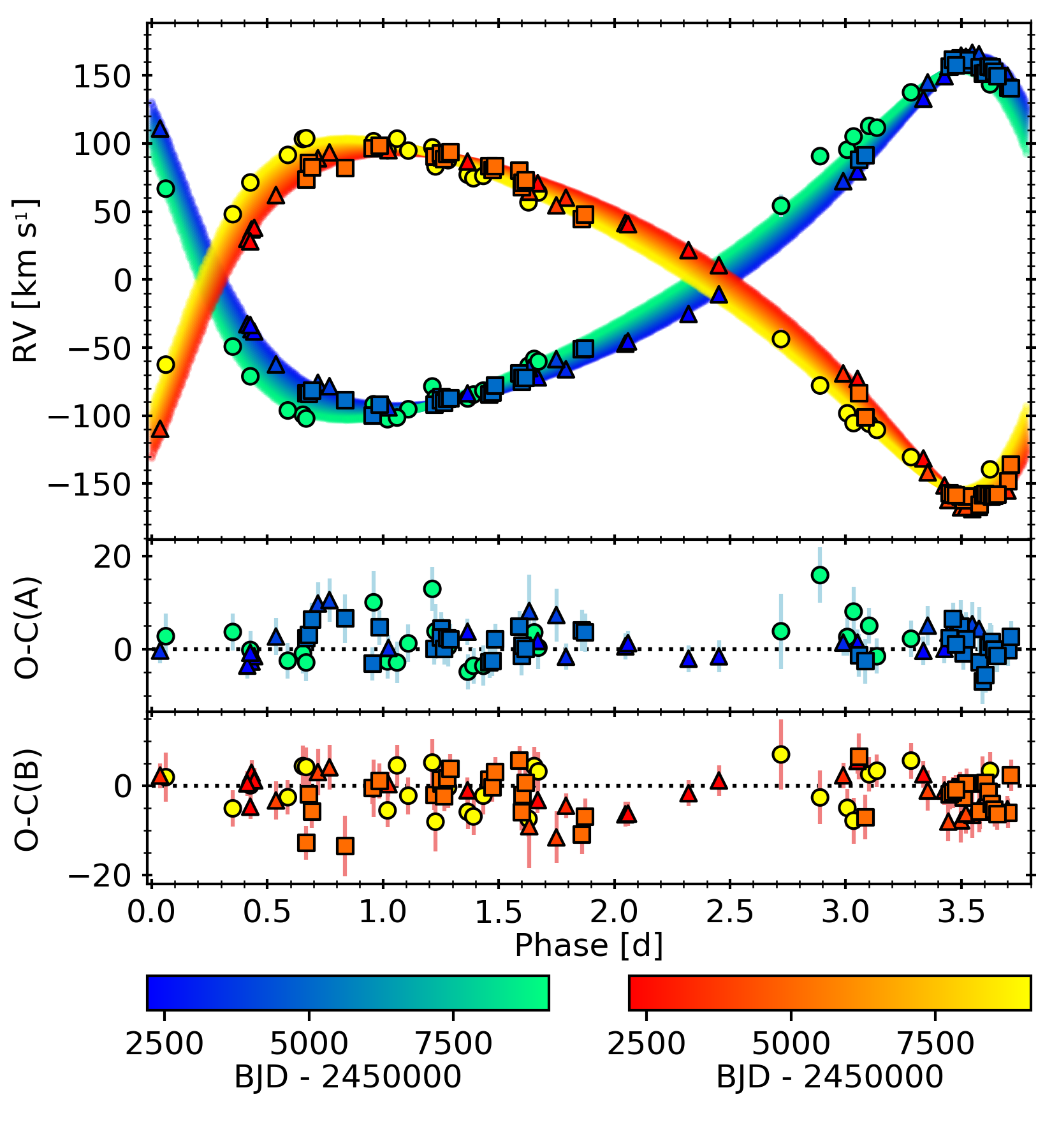}
    \caption{Best fit to the CAFE (circles), STELLA (triangles) and Tull (squares) RVs of V402\,Lac, phase folded to the anomalistic period. The argument of periastron precessing according to $\dot{\omega}$. Blue hues and red hues correspond to the primary and secondary components, respectively, while the colour code indicates the date of the measurements and fit. The bottom panels show the residuals from the best fit.}
    \label{fig:V402Lac_RV}
\end{figure}

\begin{figure*}[!t]
    \centering
    \includegraphics[width=0.9\textwidth]{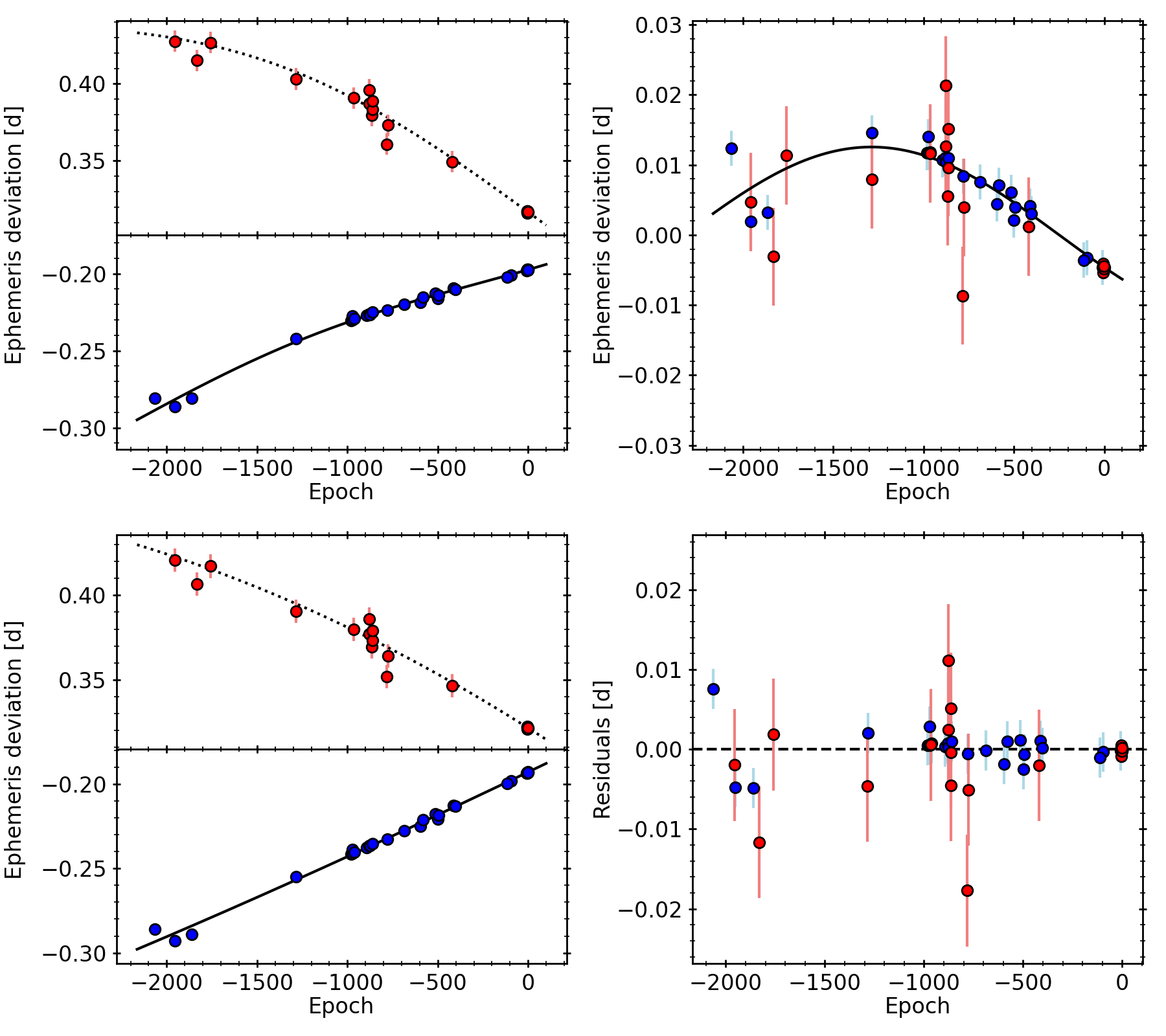}
    \caption{Same as Fig.\,\ref{fig:V889Aql_OC}, but for V402\,Lac.}
    \label{fig:V402Lac_OC}
\end{figure*}

As can be seen in Fig.\,\ref{fig:V402Lac_RV}, which shows the best fit to the RVs phase-folded to the orbital period, the relatively fast apsidal motion rate has noticeable effects over the available RVs, arising from the change of the argument of periastron with time. The upper-left panel in Fig.\,\ref{fig:V402Lac_OC} shows the best fit to the times of eclipse of V402\,Lac, with the bottom-left and upper-right panels showing the apsidal motion and third body effects separately, as in Fig.\,\ref{fig:V889Aql_OC}. While the apsidal motion effect produces a clear deviation of the eclipse timings from a linear ephemeris, the effect of a hypothetical third body is more dubious. The relatively low precision of the secondary minimum timings, the low amplitude of the modulation produced by the third body, and a period at least as long as the baseline of the observations, makes it impossible to derive precise and non-degenerate orbital parameters. For this reason, we decided to fix the eccentricity to zero, and give only a lower limit to the orbital period of the third body. Although the best fit yields an orbital period of 46 years and an amplitude of 0.015\,d, which is shown in the upper-right panel in Fig.\,\ref{fig:V402Lac_OC}, we obtain comparable fits for periods and amplitudes as low as 20 years and 0.004\,d, respectively. This lower limit to the orbital period of the third body is comparable to the period determined by \cite{Hoyman2018}, of 20.54 years, although they obtained an amplitude five times larger than the one we obtain fixing the same period. Finally, to check the significance of the third body signal and the effect that it may have over the determined orbital parameters, we repeated the analysis of the eclipse timings without including the light travel time effect in the model. We obtained a best-fit with a log-likelihood 20 units lower than the obtained with the full model, giving significant evidence in favour of the presence of a third body. Furthermore, the orbital parameters derived with the model without the effect of a third body do not differ more than one sigma from those of the full model.

\section{Light curve analysis and absolute dimensions}
\label{sec:dimensions}

As mentioned before, in this type of eclipsing binaries with similar components there is an intrinsic degeneracy between the orbital parameters and the ratio of radii. Therefore, in the following analysis of the light curves, we adopt the orbital eccentricity and the sidereal period derived in Sect.\,\ref{sec:orbital} from the study of the radial velocity curves and the eclipse timings. 

\subsection{V889\,Aql}

In order to obtain reliable photometric parameters of V889\,Aql, we have analysed the {\em TESS} photometry with the \textsc{jktebop}\footnote{\url{http://www.astro.keele.ac.uk/jkt/codes/jktebop.html}} code, a modern implementation written by J.~Southworth of the Nelson–Davis–Etzel (NDE) eclipsing binary model used in the code \textsc{ebop} \citep{Nelson1972,Etzel1981,Popper1981}. In our analysis, we adopted the sidereal period and orbital eccentricity given in Table\,\ref{tab:V889results}, and solved for the reference time of the primary eclipse ($T_{\rm I}$), the sum of the fractional radii normalised to the semi-major axis ($r_{\rm A}+r_{\rm B}$) and their ratio ($k\equiv r_{\rm B}/r_{\rm A}$), the orbital inclination ($i$), argument of periastron ($\omega$), the central surface brightness ratio of the two stars ($J$), the fraction of light from a third body divided by the total light ($L_3/L$), and a magnitude zero-point ($m_0$). 

\begin{figure}[!t]
    \centering
    \includegraphics[width=\columnwidth]{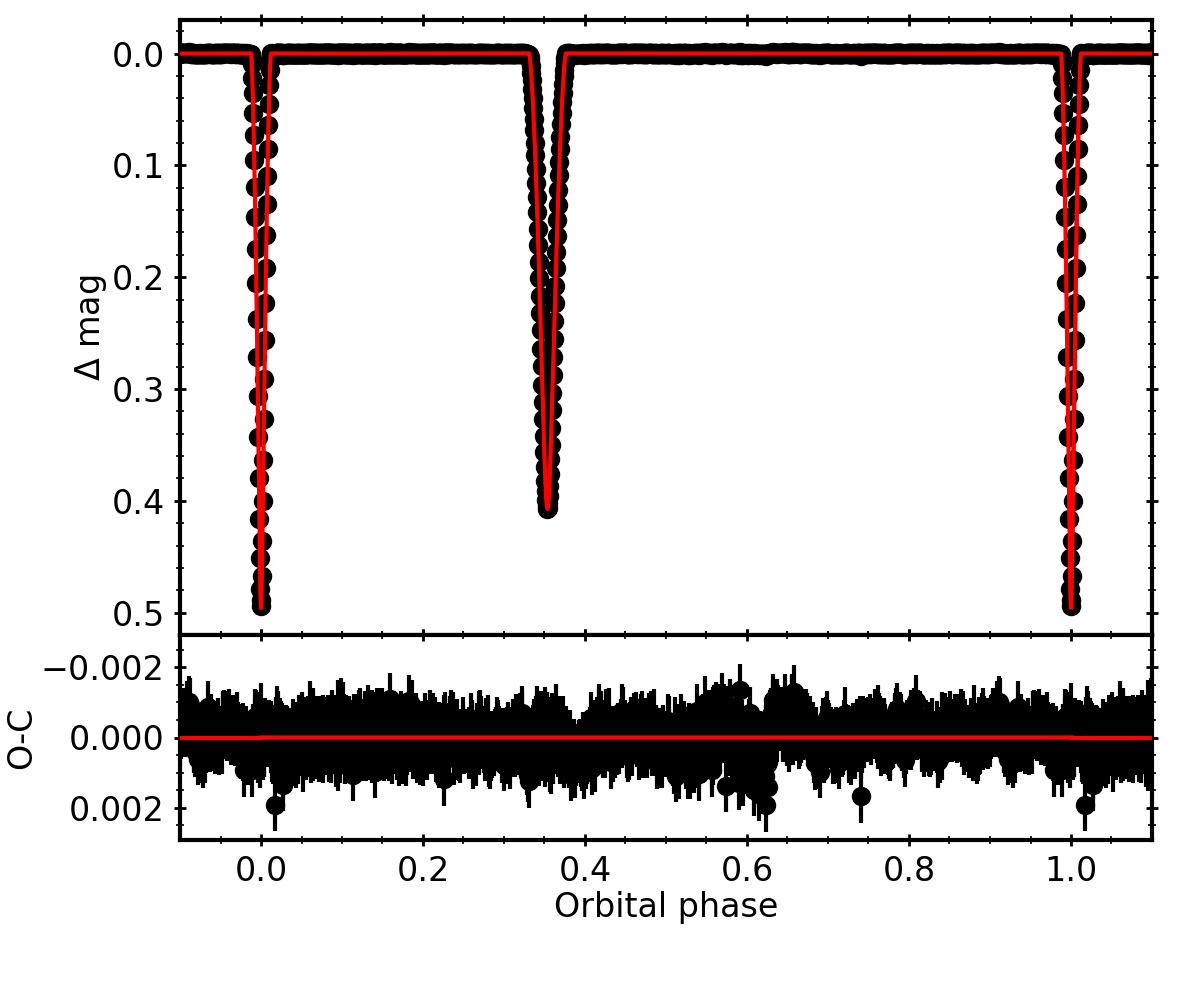}
    \includegraphics[width=\columnwidth]{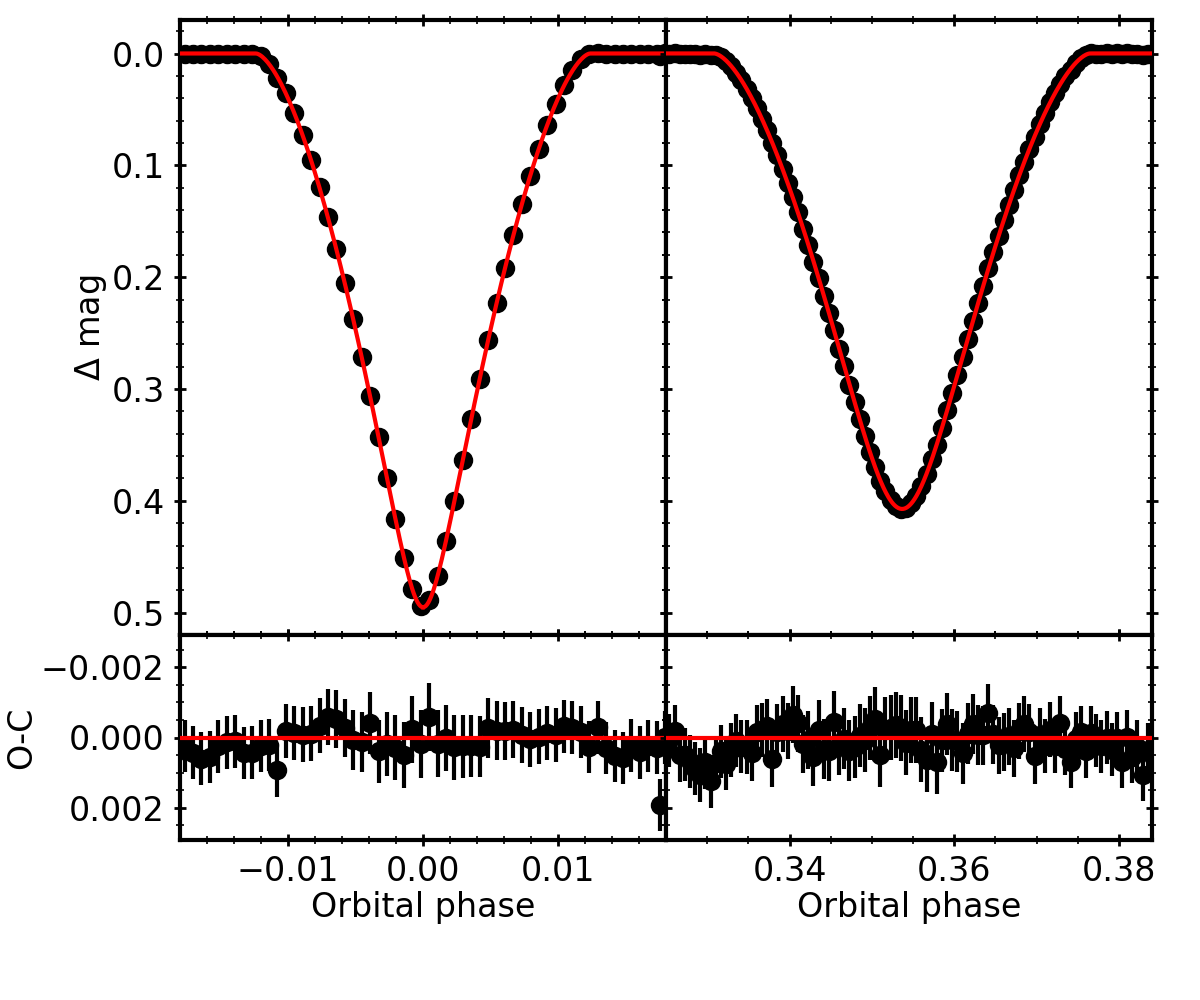}
    \caption{\textit{Top: }{\em TESS} photometry of V889\,Aql from the second orbit of Sector 40 (black dots), phase-folded to the sidereal period. The best fit computed with \textsc{jktebop} is shown as a red line. \textit{Bottom:} Details of the primary (left) and secondary (right) eclipses. The residuals of the fits are shown in the bottom part of all panels.}
    \label{fig:V889Aql_TESSfit}
\end{figure}

As shown by other authors, a well-known and common problem in the analysis of partial eclipses with similar components is that the solutions for the ratio of radii, $k$, and the surface flux ratio, $J$, are highly degenerated, and yielding solutions above and below $k=1$ \cite{Torres2015,Southworth2021}. This situation is more acute in the case of a significant amount of third light. To avoid such correlations, given that the spectroscopic analysis provided us with information about the luminosity ratio between the components, we used the resulting value, $0.98\pm0.06$, as a prior in the light curve analysis with \textsc{jktebop}. The gravity darkening and quadratic limb darkening coefficients were adopted from \cite{Claret2017}, specifically computed for the {\em TESS} wavelength range for identical components with an effective temperature of 9200\,K and $\log g = 4.20$. To compute the errors, we generated 10000 synthetic light curves with \textsc{jktebop} by adding random noise according to the observational uncertainties. We then computed the best fit and determined the standard deviation of the parameters. In addition, to determine more realistic uncertainties, we repeated the analysis fixing $e$ at the upper and lower value of the error bar in Table\,\ref{tab:V889results}, computed the dispersion of the results, and added them quadratically to the uncertainties determined with \textsc{jktebop}. This process had the highest impact over the uncertainties of $k$, $i$, and $\omega$, which were increased by a factor of $\sim10$. We show the best fit parameters and uncertainties in Table\,\ref{tab:V889TESSresults}, and show the corresponding best fit in Fig.\,\ref{fig:V889Aql_TESSfit} together with the original {\em TESS} measurements. We computed the individual fractional radii from the values of $k$ and $r_{\rm A}+r_{\rm B}$, obtaining $r_{\rm A}=0.0549\pm0.0012$ and $r_{\rm A}=0.0544\pm0.0012$. The value of the argument of periastron we find, of $126.1\pm0.2$\,deg, is in perfect agreement with the 126.14\,deg value obtained in Sect.\,\ref{sec:V889Aql}.

It should be noted that the $L_3$ value of 19.1\% confirms that found in the light curve analysis of \cite{Khaliullin1989}, who also obtained an eccentricity in perfect agreement with our value in Table\,\ref{tab:V889results}. However, our value may be overestimated due to the large pixel size of {\em TESS}, which may be including light from other close sources. Assuming a mass-luminosity relation of $L\propto M^{3.5}$, the measured third light would correspond to a third body with a mass of $\sim1.75$\,M$_{\odot}$, which is slightly higher than the minimum mass found from the light travel time effect analysis. This could suggest that the inclination of the third body orbit is close to that of the eclipsing system. However, with a third light of 19.1\% of the total light in the {\em TESS} band, we expect a third light of around 14\% in the band at which we obtained the spectra with CAFE. With this contribution to the total light, the spectral lines of the third body should be detected by analysing the spectra with \texttt{todmor}. However, we could not see any additional signal that could be attributed to the third body. This could suggest that the light from the third body is actually coming from two stars in a binary system. In this case, the outer binary system should be composed of two stars with masses $\sim$1.4\,M$_{\odot}$, each one contributing about 9.5\% to the total light in the {\em TESS} band, or $\sim5$\% in the CAFE band, which should be challenging to detect with \texttt{todmor}. In this case, the mutual inclination between the inner and the outer binary would be close to 50\,deg, thus increasing the effect that the outer system may have over the orbital parameters and apsidal motion of the inner system \cite{Borkovits2019} through the Lidov-Kozai mechanism \citep{Lidov1962,Kozai1962}. Nevertheless the time scale of the Lidov-Kozai induced oscillations, around 50000 years, would not have a significant effect in our measurement of the orbital eccentricity.

\begin{table}[!t]
\centering
\caption{Parameters of the fit to the {\em TESS} light curve of V889\,Aql.}
\label{tab:V889TESSresults}
\begin{tabular}{l c} 
\hline\hline
\noalign{\smallskip}
 Parameter & Value  \\
\noalign{\smallskip}
\hline
\noalign{\smallskip}
$P_{\rm s}$ [d] & 11.120757 (fixed) \\
\noalign{\smallskip}
$T_{\rm I}$ [BJD]  & $2459416.25979\pm0.00004$ \\
\noalign{\smallskip}
$e$ & 0.3750 (fixed) \\
\noalign{\smallskip}
$\omega$ [deg] & $126.1\pm0.2$ \\
\noalign{\smallskip}
$r_{\rm A} + r_{\rm B}$ & $0.10932\pm0.00005$ \\
\noalign{\smallskip}
$k$ & $0.99\pm0.04$ \\
\noalign{\smallskip}
$i$ [deg] & $89.06\pm0.02$ \\
\noalign{\smallskip}
$J$  & $0.971\pm0.002$ \\
\noalign{\smallskip}
$L_3/L$ & $0.191\pm0.002$ \\
\noalign{\smallskip}
$m_0$ [mag] & $0.00062\pm0.00002$ \\
\noalign{\smallskip}
\hline
\noalign{\smallskip}
\multicolumn{2}{l}{\textit{Derived parameters}} \\
\noalign{\smallskip}
\hline
\noalign{\smallskip}
$r_{\rm A}$ & $0.0549\pm0.0012$ \\
\noalign{\smallskip}
$r_{\rm B}$ & $0.0544\pm0.0012$ \\
\noalign{\smallskip}
\hline
\end{tabular}
\end{table}

Using the results presented in Tables\,\ref{tab:V889results} and \ref{tab:V889TESSresults}, the absolute dimensions of the eclipsing components of V889\,Aql are given in Table\,\ref{tab:absdimensions}. 

\subsection{V402\,Lac} \label{sec:V402LacLC}

As in the case of V889\,Aql, we have modelled the {\em TESS} light curves of V402\,Lac adopting the sidereal period and the orbital eccentricity from the study in Section\,\ref{sec:V402Lac}. Given the larger stellar deformation of the components of V402\,Lac, apparent in the out-of-eclipse photometry, and the relatively long duration of the eclipses, we used the same \textsc{jktebop} code only for preliminary solutions, while the final parameters were computed using the Wilson-Devinney \citep[hereafter \textsc{WD},][]{WD1971} code instead. This code represents the stars using Roche geometry and takes into account reflection and proximity effects accordingly. We fixed the effective temperature to the values indicated in Table\,\ref{tab:todmorV402}, and adopted the quadratic limb-darkening coefficients by \cite{Claret2017} corresponding to these temperatures. We also adopted the mass ratio from the spectroscopic solution, and fixed the bolometric albedo and the gravity darkening exponent to 1, as usual for radiative envelopes. The rotation of the components was assumed to be pseudo-synchronised at periastron \citep{Hut1981}, although we also checked that the results obtained assuming the broadening values listed in Table\,\ref{tab:todmorV402} did not change significantly. We solved for the time of primary eclipse, the argument of periastron, the orbital inclination, the gravitational potentials of each component ($\Omega_{\rm A}$ and $\Omega_{\rm B}$), the luminosity of the component A (as a parameter depending on the light curve normalisation), and the third light contribution. Unlike for V889\,Aql, we could not fix the value of $L_{\rm B}/L_{\rm A}$ from the spectroscopic analysis, since we could not find a precise value different from 1. Therefore, we opted for constraining its value by fixing the effective temperature of star B and assuming radiative coupling between the components for the $I$-band in the WD code. $L_{\rm B}/L_{\rm A}$ is finally reported in the table, and justifies the equal-luminosity assumption we made in the spectroscopic analysis.

\begin{figure}[!t]
    \centering
    \includegraphics[width=\columnwidth]{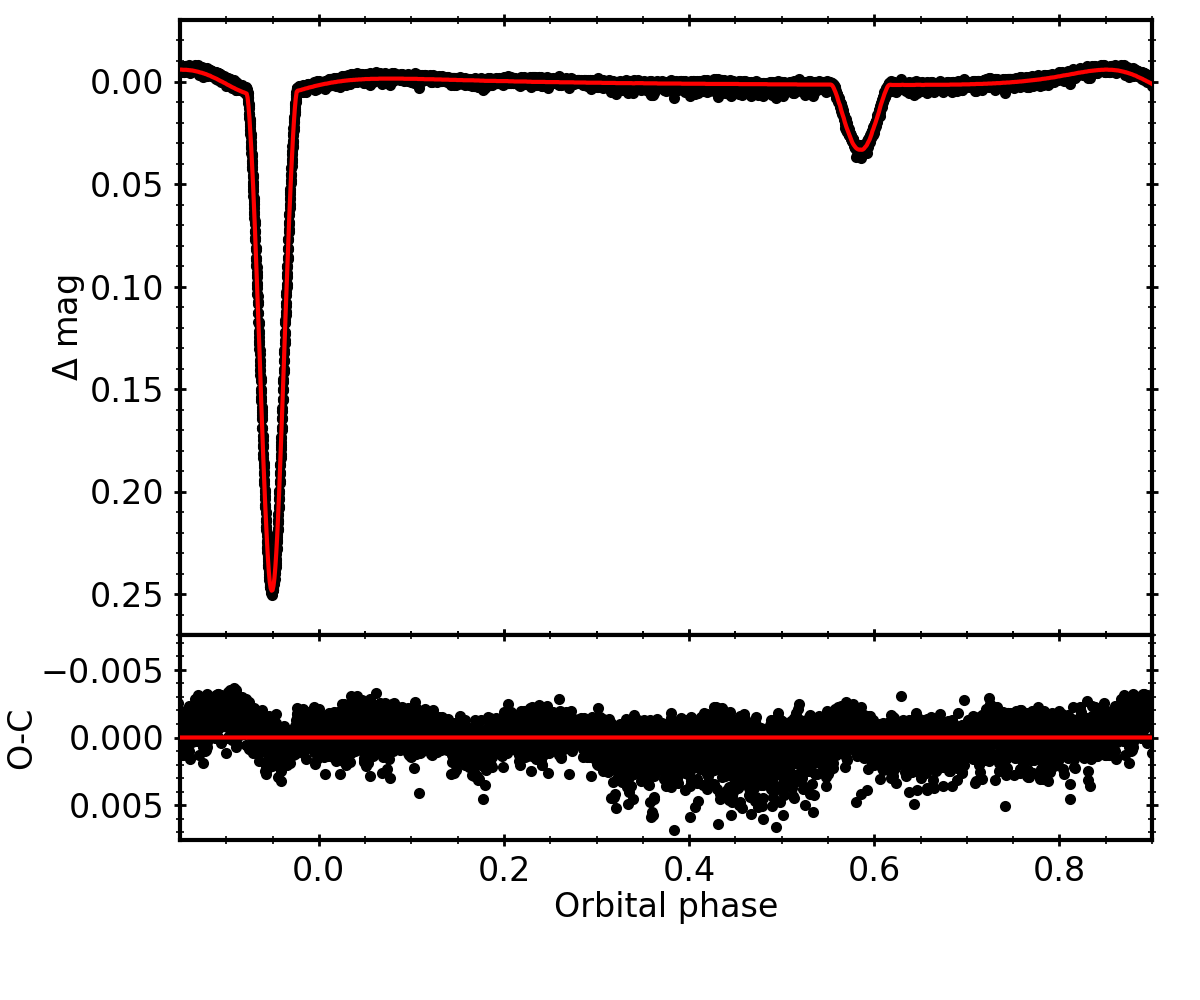}
    \includegraphics[width=\columnwidth]{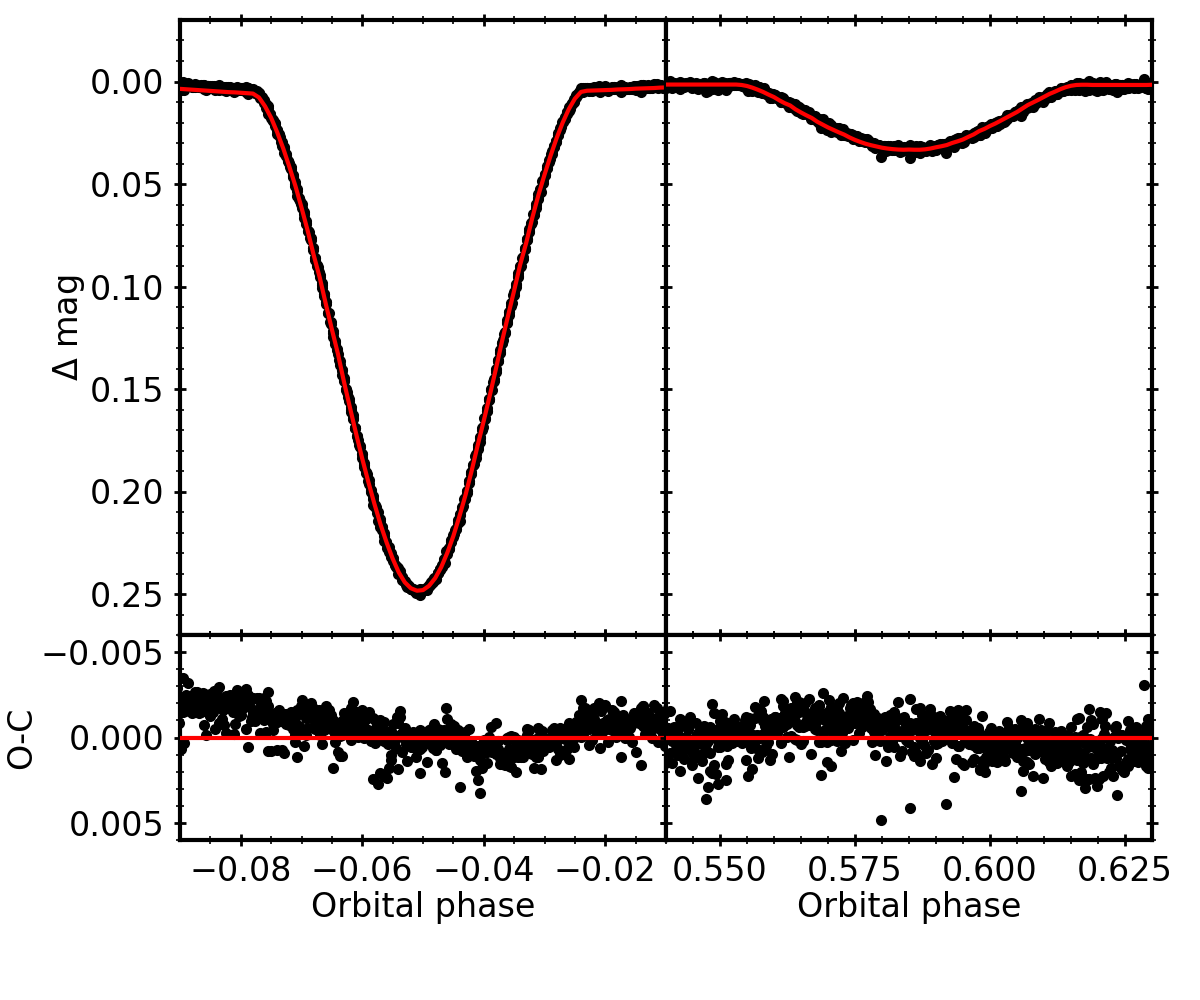}
    \caption{\textit{Top: }{\em TESS} photometry of V402\,Lac from the second orbit of Sector 16 (black dots), phase-folded to the sidereal period. The best fit computed with the \textsc{WD} code is shown as a red line. \textit{Bottom:} Details of the primary (left) and secondary (eclipses). The residuals of the fits are shown in the bottom part of all panels.}
    \label{fig:V402Lac_TESSfit}
\end{figure}

We evaluated the uncertainties following the same process done for V889\,Aql. We reanalysed the light curve with the eccentricity fixed at the value found in the spectroscopic solution and at the upper and lower values of the error bar in Table\,\ref{tab:V889results}, computed the dispersion of the solutions, and added them quadratically to the formal uncertainties determined with the \textsc{WD} code. We obtained errors much larger than the formal uncertainties computed by the \textsc{WD} code, as expected. We list in Table\,\ref{tab:V402TESSresults} the parameters of the best solution and the corresponding uncertainties, and shown in Fig.\,\ref{fig:V402Lac_TESSfit} the best fit to the data. 

As can be seen in the residuals of the best fit, there is a systematic deviation of the model at the beginning of the primary eclipse. This systematic effect is repeated when using the two parts of the {\em TESS} light curve individually, which rules out that it is produced by trends in the data processing. We attribute this systematic deviation to a lack of accuracy in the \textsc{WD} code in modeling the stellar deformations producing the reflection effect. It should be mentioned that the phase of periastron coincides with the beginning of the primary eclipse in the light curve of V402 Lac. Therefore, to prevent the underestimation of the errors due to any bias that may arise from this systematic deviation, we inflated the errors by a factor of two. In any case, all solutions agree with a system of almost identical components.

We obtained an argument of periastron of $58.1\pm1.4$\,d, fully compatible with the value corresponding to the same reference time derived by the joint fit to RVs and times of minima. From the values of the gravitational potential we derive individual fractional radii of $r_{\rm A}=0.1315\pm0.0018$ and $r_{\rm B}=0.1304\pm0.0014$. The luminosity ratio as derived with the \textsc{WD} code is $0.99\pm0.02$, which is compatible with the value of 1 that we fixed in the spectroscopic analysis with \texttt{todmor}. In addition, we detect a significant third light contribution of $15\pm4$\%. Although some of this light may be coming from a third companion, as seen in by the mask used by {\em TESS} shown in Fig.\,\ref{fig:V402mask}, a significant part of it might have its origin in the star (label \#2) inside the {\em TESS} mask and in the bright star (label \#3) close to V402\,Lac. 

\begin{table}[!t]
\centering
\caption{Parameters of the fit to the {\em TESS} light curve of V402\,Lac.}
\label{tab:V402TESSresults}
\begin{tabular}{l c} 
\hline\hline
\noalign{\smallskip}
 Parameter & Value  \\
\noalign{\smallskip}
\hline
\noalign{\smallskip}
$P_{\rm s}$ [d] & 3.782043 (fixed) \\
\noalign{\smallskip}
$T_{\rm I}$ [BJD]  &  2458761.7191$\pm$0.0014 \\
\noalign{\smallskip}
$e$ & 0.376 (fixed) \\
\noalign{\smallskip}
$\omega$ [deg] & $58.1\pm1.4$ \\
\noalign{\smallskip}
$i$ [deg] & $79.99\pm0.12$ \\
\noalign{\smallskip}
$L_{\rm B}/L_{\rm A}$ & $0.99\pm0.02$  \\
\noalign{\smallskip}
$L_3/L$ & $0.15\pm0.04$ \\
\noalign{\smallskip}
$\Omega_{\rm A}$ & $9.28\pm0.10$ \\
\noalign{\smallskip}
$\Omega_{\rm B}$ &  $9.29\pm0.04$\\
\noalign{\smallskip}

\hline
\noalign{\smallskip}
\multicolumn{2}{l}{\textit{Derived parameters}} \\
\noalign{\smallskip}
\hline
\noalign{\smallskip}
$r_{\rm A}$ & $0.1315\pm0.0018$ \\
\noalign{\smallskip}
$r_{\rm B}$ & $0.1304\pm0.0014$ \\
\noalign{\smallskip}
\hline
\end{tabular}
\end{table}

\begin{table}[!t]
\centering
\caption{Absolute dimensions of V889\,Aql and V402\,Lac.}
\label{tab:absdimensions}
\begin{tabular}{l c c} 
\hline\hline
\noalign{\smallskip}
 & V889\,Aql & V402\,Lac  \\
\noalign{\smallskip}
\hline
\noalign{\smallskip}
$M_{\rm A}$ [M$_{\odot}$] & $2.17\pm0.02$  & $2.80\pm0.05$ \\
\noalign{\smallskip}
$M_{\rm B}$ [M$_{\odot}$] & $2.13\pm0.02$  &  $2.78\pm0.05$ \\
\noalign{\smallskip}
$R_{\rm A}$ [R$_{\odot}$] & $1.87\pm0.04$  & $2.38\pm0.03$ \\
\noalign{\smallskip}
$R_{\rm B}$ [R$_{\odot}$] & $1.85\pm0.04$  & $2.36\pm0.03$ \\
\noalign{\smallskip}
$\log g_{\rm A}$ & $4.23\pm0.02$  & $4.132\pm0.013$ \\
\noalign{\smallskip}
$\log g_{\rm B}$ & $4.23\pm0.02$ &  $4.136\pm0.014$ \\
\noalign{\smallskip}
\hline
\noalign{\smallskip}
\multicolumn{3}{l}{\textit{Derived from models}}\\
\noalign{\smallskip}
\hline
\noalign{\smallskip}
$\log k_{\rm 2,A}$ & $-2.40\pm0.02$ & $-2.40\pm0.02$  \\
\noalign{\smallskip}
$\log k_{\rm 2,B}$ & $-2.41\pm0.02$ & $-2.40\pm0.02$  \\
\noalign{\smallskip}
$T_{\rm eff,A}$ [K] & $9210\pm110$ & $11330\pm250$  \\
\noalign{\smallskip}
$T_{\rm eff,B}$ [K] & $9075\pm110$ & $11280\pm240$   \\
\noalign{\smallskip}
$t_{\rm A}$ [Ma] & $200\pm40$  & $230\pm20$  \\
\noalign{\smallskip}
$t_{\rm B}$ [Ma] & $210\pm40$ & $230\pm20$   \\
\noalign{\smallskip}
$L_{\rm bol,A}$ [L$_{\odot}$] & $22.6\pm1.4$  & $85\pm8$  \\
\noalign{\smallskip}
$L_{\rm bol,B}$ [L$_{\odot}$] & $20.9\pm1.4$ & $82\pm7$   \\
\noalign{\smallskip}
\hline
\end{tabular}
\end{table}

\section{Discussion and conclusions}
\label{sec:conclusions}

Absolute dimensions of the components of V889\,Aql and V402\,Lac, as derived from the above orbital elements and light curve solutions, are given in Table\,\ref{tab:absdimensions}, with relative errors between 0.8 and 2.1\,\%. These absolute dimensions are sufficiently precise for a meaningful comparison with theoretical models and the analysis of the observed apsidal motion rates. Nevertheless, a reliable comparison with the distance given by {\em Gaia}, as shown in Table\,\ref{tab:proprties}, was not possible due to the uncertainties in the contribution of the third light components measured with {\em TESS} in the $V$-band photometry and the computation of the extinction coefficients, expected to be relatively large. We could only verify that a reasonable combination of adopted third light and extinction could reproduce the {\em Gaia} measurements.

\begin{figure}[!t]
    \centering
    \includegraphics[width=\columnwidth]{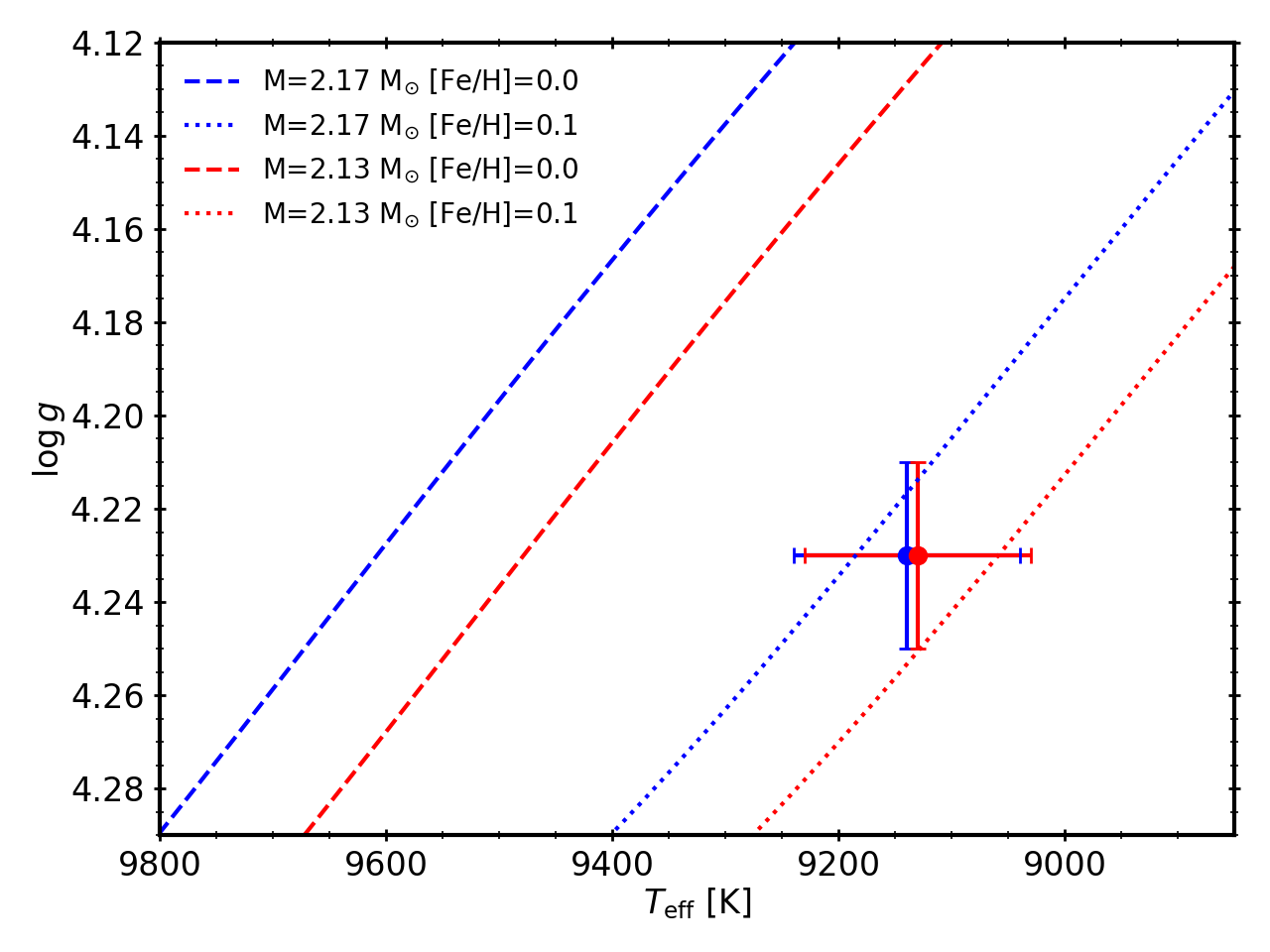}
    \caption{$T_{\rm eff}$-$\log g$ relationship for the primary (blue) and secondary (red) components of V889\,Aql, together with the theoretical models from \cite{Claret2019} corresponding to the mass of each component and metallicities of 0.0 (dashed lines) and 0.1 (dotted lines).}
    \label{fig:V889model}
\end{figure}

\begin{figure}[!t]
    \centering
    \includegraphics[width=\columnwidth]{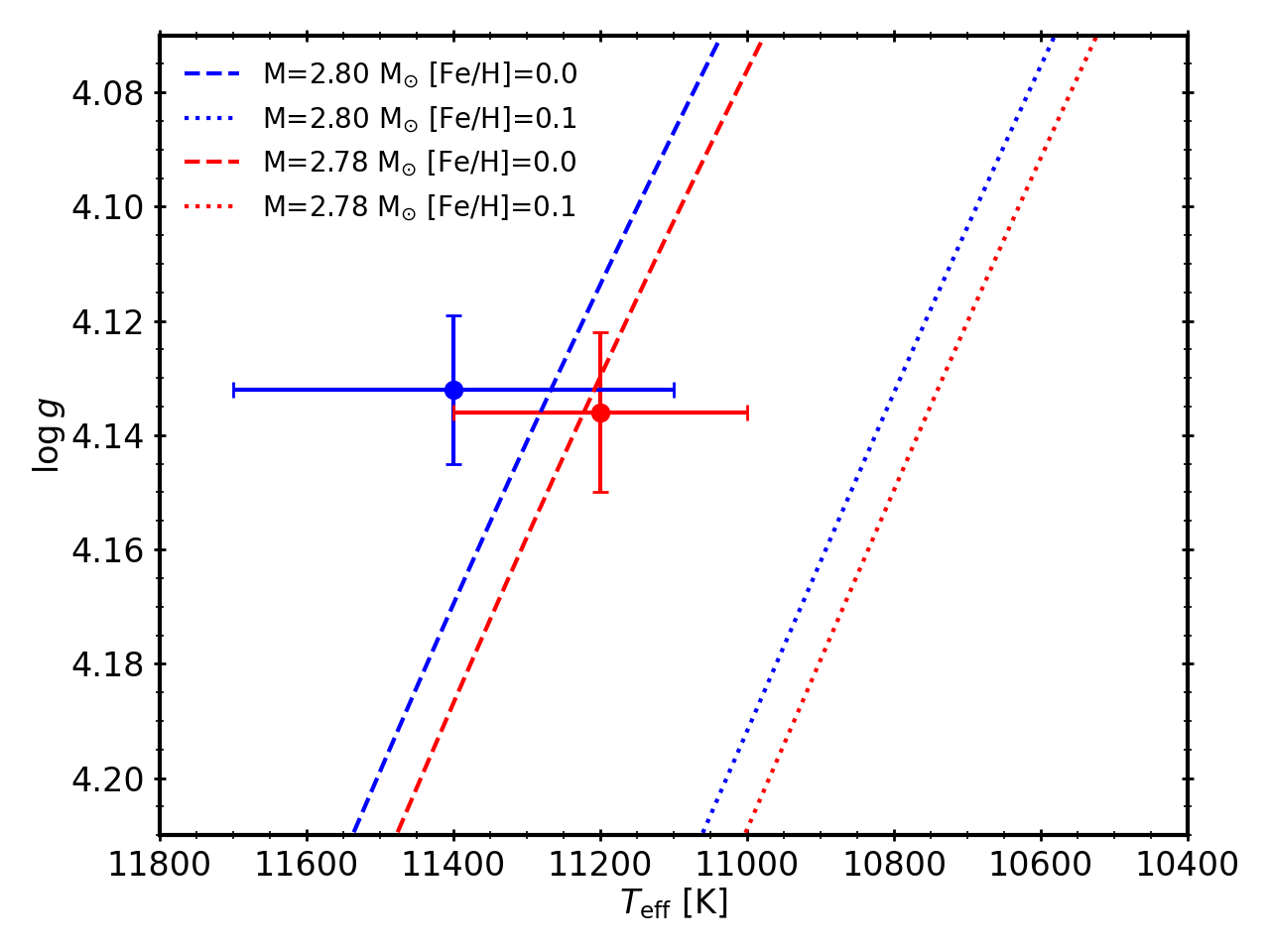}
    \caption{$T_{\rm eff}$-$\log g$ relationship for the primary (blue) and secondary (red) components of V402\,Lac, together with the theoretical models from \cite{Claret2019} corresponding to the mass of each component and metallicities of 0.0 (dashed lines) and 0.1 (dotted lines).}
    \label{fig:V402model}
\end{figure}

We computed the theoretical values of the $k_2$ using the theoretical models published by \cite{Claret2019}, which are based on the Modules for Experiments in Stellar Astrophysics package \citep[MESA;][]{Paxton2011,Paxton2013,Paxton2015}. These models assume convective core overshooting following the semi-empirical dependence on mass given by \cite{Claret2018}. We homogeneously derived $k_2$ values and their errors from interpolation in the models, using $M$ and $\log g$ as input parameters. However, since there are no previous determinations of the metallicity of any of the two systems, we tuned them until the derived mean effective temperature derived was compatible with the mean effective temperatures listed in Tables\,\ref{tab:todmorV889} and \ref{tab:todmorV402}. This was achieved by assuming a metallicity of [Fe/H]$=+0.10\pm0.02$ for V889\,Aql and [Fe/H]$= 0.00\pm0.05$ for V402\,Lac, as can be seen in Figs. \ref{fig:V889model} and \ref{fig:V402model}. Additional cross-checks were carried out by comparing the derived ages of the components two components, obtaining compatible values in both systems, thus providing additional reassurance. The uncertainties of the theoretical values were computed from the standard deviations of the results obtained from $10^5$ random resamplings of the input parameters, $M$, $\log g$, and $[Fe/H]$, employing their quoted uncertainties.

Using the absolute dimensions and orbital parameters of V889\,Aql, we obtain a theoretical apsidal motion rate of $\dot{\omega}=0.000426\pm0.000007$\,deg\,cycle$^{-1}$, with a classical term of $\dot{\omega}=0.000075\pm0.000007$\,deg\,cycle$^{-1}$ and a relativistic term of $\dot{\omega}=0.000336\pm0.000002$\,deg\,cycle$^{-1}$. Additionally, the third body companion, hypothetically composed of two stars of 1.4\,M$_{\odot}$ and with a mutual inclination of 50\,deg, contributes with an estimated value of $\dot{\omega}=0.000030$\,deg\,cycle$^{-1}$ \citep{Borkovits2019}, which increases the theoretical apsidal motion value to $\dot{\omega}=0.00046$\,deg\,cycle$^{-1}$. This value agrees well with the observed value of $\dot{\omega}=0.00046\pm0.00002$\,deg\,cycle$^{-1}$, although the assumption on the third body mass and inclination remains a source of significant uncertainty.

Using the absolute dimensions and orbital parameters of V402\,Lac, we obtain a theoretical apsidal motion of $\dot{\omega}=0.0088\pm0.0005$\,deg\,cycle$^{-1}$, with a classical term of $\dot{\omega}=0.0080\pm0.0005$\,deg\,cycle$^{-1}$ and a relativistic term of $\dot{\omega}=0.000822\pm0.000007$\,deg\,cycle$^{-1}$. Despite not considering the effect of the potential third body, which is nevertheless not expected to be significant given the large classical term, the predicted value is compatible with the observed $\dot{\omega}=0.0090\pm0.0003$\,deg\,cycle$^{-1}$ within errors.

In conclusion, both systems are now part of the select group of eclipsing binaries with precise absolute dimensions, necessary for the improvement of empirical calibrations \citep{Torres2010} or the test of apsidal rates \citep{Baroch2021}. The case of V889\,Aql is a clear example of apsidal motion dominated by the relativistic term, 75\% of the total, with a significant third body contribution. On the other hand, V402\,Lac shows a rather fast apsidal motion dominated by the classical term, with negligible relativistic and third body contributions. In the light curves, significant third light contributions were identified as well as the presence of a third body through the variations of the eclipse timings, definitely in the case of V889\,Aql.

\begin{acknowledgements}
This article has made use of observations collected at the Centro Astron\'omico Hispano Alem\'an (CAHA) at Calar Alto, operated jointly by the Junta de Andaluc\'ia and the Instituto de Astrof\'isica de Andaluc\'ia (CSIC). Data were partly obtained with the TIGRE telescope, located at La Luz observatory, México. TIGRE is a collaboration of the Hamburger Sternwarte, the Universities of Hamburg, Guanajuato and Liège. This paper includes data collected by the {\em TESS} mission. Funding for the {\em TESS} mission is provided by the NASA's Science Mission Directorate. STELLA was made possible by funding through the State of Brandenburg (MWFK) and the German Federal Ministry of Education and Research (BMBF). The facility is a collaboration of the AIP in Brandenburg with the IAC in Tenerife. The construction of the 2.7\,m telescope and Tull spectrograph was supported by contract NASr-242 with the National Aeronautics and Space Administration. This work has made use of data from the European Space Agency (ESA) mission {\it Gaia}. {\em Gaia} data are being processed by the {\em Gaia} Data Processing and Analysis Consortium (DPAC). Funding for the DPAC is provided by national institutions, in particular the institutions participating in the {\em Gaia} MultiLateral Agreement (MLA). The {\em Gaia} mission website is \url{https://www.cosmos.esa.int/gaia}. The {\em Gaia} archive website is \url{https://archives.esac.esa.int/gaia}. This publication has been made possible by grant PGC2018-098153-B-C33 funded by MCIN/AEI/10.13039/501100011033 and by “ERDF A way of making Europe”, by the program Unidad de Excelencia María de Maeztu CEX2020-001058-M, and by the Generalitat de Catalunya/CERCA programme. 

\end{acknowledgements}

\bibliographystyle{aa} 
\bibliography{bibtex.bib} 

\begin{appendix}
\section{RV data}
Tables \ref{tab:V889AqlRVs} and \ref{tab:V402LacRVs} list the RVs used in this work, as computed with \texttt{todmor}. The RVs from each instrument are labelled as C (CAFE), T (TIGRE), Tu (Tull), and S (STELLA).

\begin{table}[h]
\centering
\caption{Radial velocities of V889\,Aql.}
\label{tab:V889AqlRVs}
\begin{tabular}{l ccc} 
\hline\hline
\noalign{\smallskip}
 BJD & RV$_{\rm A}$ & RV$_{\rm B}$ & Ins. \\
     & [km\,s$^{-1}$] & [km\,s$^{-1}$] & \\
\hline
\noalign{{\smallskip}}
2459102.3675 & $42.469\pm0.325$ & $-85.62\pm0.42$ & C  \\ 
2459103.3310 & $40.454\pm0.301$ & $-83.30\pm0.39$ & C  \\ 
2459103.3826 & $39.666\pm0.278$ & $-82.56\pm0.37$ & C  \\ 
2459114.3370 & $41.624\pm0.319$ & $-84.50\pm0.43$ & C  \\ 
2459114.3744 & $41.201\pm0.445$ & $-84.13\pm0.56$ & C  \\ 
2459115.3202 & $14.186\pm0.394$ & $-55.87\pm0.53$ & C  \\ 
2459116.3237 & $-71.968\pm0.366$ & $30.90\pm0.50$ & C  \\ 
2459116.3702 & $-76.775\pm0.903$ & $36.40\pm1.33$ & C  \\ 
2459121.3077 & $-5.061\pm0.298$ & $-37.28\pm0.38$ & C  \\ 
2459121.3253 & $-4.744\pm0.293$ & $-37.76\pm0.38$ & C  \\ 
2459122.2978 & $14.273\pm0.299$ & $-56.48\pm0.40$ & C  \\ 
2459122.3335 & $15.184\pm0.351$ & $-56.34\pm0.47$ & C  \\ 
2459132.3424 & $-6.924\pm0.354$ & $-36.18\pm0.47$ & C  \\ 
2459134.3353 & $28.252\pm0.365$ & $-71.86\pm0.50$ & C  \\ 
2459134.3504 & $28.721\pm0.408$ & $-72.33\pm0.54$ & C  \\ 
2459135.3316 & $39.934\pm0.317$ & $-83.20\pm0.42$ & C  \\ 
2459135.3644 & $39.968\pm0.312$ & $-83.53\pm0.41$ & C  \\ 
2459146.2947 & $38.837\pm0.886$ & $-81.91\pm1.16$ & C  \\ 
2459148.2882 & $31.171\pm0.615$ & $-72.87\pm0.80$ & C  \\ 
2459300.9837 & $29.097\pm0.819$ & $-65.96\pm1.03$ & T  \\ 
2459303.9885 & $32.721\pm0.756$ & $-69.75\pm1.02$ & T  \\ 
2459307.9894 & $-66.936\pm0.583$ & $33.51\pm0.80$ & T  \\ 
2459309.9585 & $-11.892\pm0.891$ & $-25.16\pm1.24$ & T  \\ 
2459311.9485 & $25.763\pm0.606$ & $-64.17\pm0.77$ & T  \\ 
2459313.9349 & $44.741\pm0.615$ & $-84.86\pm0.87$ & T  \\ 
2459315.9387 & $-13.346\pm0.820$ & $-23.41\pm1.14$ & T  \\ 
2459317.9266 & $-110.733\pm0.685$ & $76.74\pm0.90$ & T  \\ 
2459319.9227 & $-42.241\pm0.638$ & $8.53\pm0.90$ & T  \\ 
2459321.9157 & $7.355\pm0.707$ & $-44.38\pm0.85$ & T  \\ 
2459322.9755 & $25.088\pm0.678$ & $-62.02\pm0.93$ & T  \\ 
2459323.9769 & $39.146\pm0.575$ & $-76.52\pm0.84$ & T  \\ 
2459325.9019 & $40.406\pm0.617$ & $-77.65\pm0.89$ & T  \\ 
2459327.9296 & $-97.977\pm0.828$ & $62.12\pm1.06$ & T  \\ 
2459329.9394 & $-78.204\pm0.856$ & $40.71\pm1.20$ & T  \\ 
\noalign{{\smallskip}}
\hline
\end{tabular}
\end{table}

\begin{table}[!t]
\centering
\caption{Radial velocities of V402\,Lac.}
\label{tab:V402LacRVs}
\begin{tabular}{l ccc} 
\hline\hline
\noalign{\smallskip}
 BJD & RV$_{\rm A}$ & RV$_{\rm B}$ & Ins. \\
     & [km\,s$^{-1}$] & [km\,s$^{-1}$] & \\
\hline
\noalign{{\smallskip}}
2452205.6795 & $-23.2\pm1.4$ & $23.6\pm1.4$ & Tu  \\ 
2452206.6948 & $134.9\pm1.7$ & $-129.4\pm1.7$ & Tu  \\ 
2452206.7861 & $151.5\pm1.8$ & $-149.3\pm1.9$ & Tu  \\ 
2452207.5727 & $-34.4\pm1.3$ & $38.8\pm1.3$ & Tu  \\ 
2452207.5842 & $-36.1\pm1.3$ & $40.0\pm1.3$ & Tu  \\ 
2452208.8099 & $-69.8\pm1.3$ & $72.8\pm1.4$ & Tu  \\ 
2452209.5928 & $-8.8\pm2.3$ & $12.5\pm2.3$ & Tu  \\ 
2452254.5742 & $-45.1\pm1.1$ & $43.5\pm1.1$ & Tu  \\ 
2452254.5857 & $-43.4\pm1.0$ & $42.7\pm1.0$ & Tu \\ 
2452255.5780 & $81.5\pm1.2$ & $-70.9\pm1.1$ & Tu  \\ 
2452256.7221 & $-30.8\pm1.2$ & $31.9\pm1.2$ & Tu \\ 
2452256.7350 & $-31.4\pm1.1$ & $30.1\pm1.2$ & Tu \\ 
2452257.5633 & $-84.5\pm1.1$ & $91.7\pm1.1$ & Tu \\ 
2452257.6743 & $-81.6\pm1.2$ & $88.8\pm1.3$ & Tu \\ 
2453218.7628 & $-63.9\pm1.2$ & $62.2\pm1.2$ & Tu \\ 
2453307.6607 & $151.7\pm1.3$ & $-153.0\pm1.4$ & Tu  \\ 
2453307.7799 & $113.0\pm1.1$ & $-107.6\pm1.2$ & Tu \\ 
2453308.7670 & $-91.9\pm1.0$ & $97.0\pm1.1$ & Tu \\ 
2453310.7328 & $74.3\pm1.3$ & $-66.9\pm1.3$ & Tu \\ 
2453889.9096 & $166.6\pm3.9$ & $-165.4\pm4.3$ & Tu  \\ 
2453889.9579 & $168.6\pm4.0$ & $-166.6\pm4.5$ & Tu \\ 
2453890.9115 & $-73.8\pm3.8$ & $91.1\pm4.5$ & Tu \\ 
2453890.9613 & $-76.4\pm3.8$ & $95.5\pm4.4$ & Tu \\ 
2453891.8249 & $-63.0\pm7.4$ & $66.5\pm9.0$ & Tu \\ 
2453891.9420 & $-56.4\pm5.1$ & $56.6\pm5.2$ & Tu \\ 
2453972.9724 & $146.9\pm3.4$ & $-139.7\pm3.7$ & Tu  \\ 
2453973.9370 & $-60.3\pm3.0$ & $64.1\pm3.5$ & Tu \\ 
2453976.8437 & $158.6\pm3.5$ & $-160.0\pm3.5$ & Tu  \\ 
2453976.9196 & $165.7\pm3.4$ & $-165.2\pm3.6$ & Tu \\ 
2453976.9762 & $167.6\pm3.9$ & $-165.0\pm5.0$ & Tu \\ 
2455077.4986 & $144.8\pm1.7$ & $-177.2\pm1.8$ & S \\ 
2455077.5114 & $139.8\pm1.9$ & $-180.3\pm1.7$ & S  \\ 
2455077.5241 & $143.4\pm1.7$ & $-177.6\pm1.7$ & S  \\ 
2455078.7390 & $-117.9\pm2.1$ & $78.7\pm2.1$ & S  \\ 
2455079.3836 & $-92.8\pm3.0$ & $49.9\pm2.8$ & S  \\ 
2455079.6449 & $-69.1\pm3.3$ & $26.5\pm3.2$ & S  \\ 
2455079.6576 & $-68.5\pm2.9$ & $29.6\pm2.7$ & S  \\ 
2455081.4869 & $123.0\pm1.6$ & $-166.0\pm1.5$ & S  \\ 
2455081.4997 & $122.5\pm1.6$ & $-154.1\pm1.9$ & S  \\ 
2455082.5519 & $-110.1\pm2.2$ & $80.3\pm2.6$ & S  \\ 
2455139.5204 & $-109.9\pm1.6$ & $72.5\pm1.5$ & S  \\ 
2455139.5514 & $-104.5\pm1.8$ & $73.9\pm1.7$ & S  \\ 
2455143.3322 & $-106.4\pm1.4$ & $74.8\pm1.5$ & S  \\ 
2455143.3449 & $-108.5\pm1.5$ & $70.7\pm1.7$ & S  \\ 
2455143.3576 & $-105.6\pm1.7$ & $74.1\pm2.0$ & S  \\ 
2455143.3712 & $-105.4\pm1.6$ & $75.8\pm1.7$ & S  \\ 
2455147.3229 & $-102.3\pm1.5$ & $65.2\pm1.5$ & S  \\ 
2455147.3357 & $-101.2\pm1.4$ & $62.8\pm1.3$ & S  \\ 
2455147.3484 & $-95.8\pm1.6$ & $65.6\pm1.5$ & S  \\ 
2455147.4530 & $-87.1\pm1.6$ & $61.9\pm1.4$ & S  \\ 
2455147.4657 & $-90.4\pm1.6$ & $53.4\pm1.5$ & S  \\ 
2455147.4787 & $-90.2\pm1.5$ & $55.3\pm1.4$ & S  \\ 
2455149.3138 & $138.5\pm1.6$ & $-175.0\pm1.5$ & S  \\ 
2455149.3266 & $143.9\pm2.0$ & $-175.8\pm1.8$ & S  \\ 
2455149.3393 & $139.6\pm1.7$ & $-176.3\pm1.6$ & S  \\ 
2455149.4421 & $138.1\pm2.7$ & $-183.7\pm2.5$ & S  \\ 
2455149.4549 & $133.3\pm3.8$ & $-176.6\pm5.3$ & S  \\ 
2455149.4677 & $133.8\pm2.2$ & $-175.5\pm2.3$ & S  \\ 
2455149.4805 & $138.7\pm2.2$ & $-176.3\pm2.4$ & S  \\ 
2455149.4933 & $138.2\pm2.1$ & $-177.5\pm2.0$ & S  \\ 
2455149.5060 & $134.9\pm2.2$ & $-177.1\pm2.4$ & S  \\ 
2455149.5190 & $131.4\pm1.9$ & $-176.0\pm1.9$ & S  \\ 
2455150.3125 & $-101.5\pm2.3$ & $55.6\pm2.4$ & S  \\ 
2455150.3252 & $-101.9\pm3.9$ & $67.6\pm3.8$ & S  \\ 
\noalign{{\smallskip}}
\hline
\end{tabular}
\end{table}

\addtocounter{table}{-1}
\begin{table}[!t]
\centering
\caption{Continued.}
\begin{tabular}{l ccc} 
\hline\hline
\noalign{\smallskip}
 BJD & RV$_{\rm A}$ & RV$_{\rm B}$ & Ins. \\
     & [km\,s$^{-1}$] & [km\,s$^{-1}$] & \\
\hline
\noalign{{\smallskip}}
2455150.3379 & $-99.6\pm2.0$ & $64.6\pm2.1$ & S  \\ 
2455150.4817 & $-106.6\pm4.3$ & $64.2\pm6.1$ & S  \\ 
2455156.4852 & $70.0\pm3.5$ & $-101.2\pm4.3$ & S  \\ 
2455156.5117 & $73.4\pm3.8$ & $-119.5\pm4.0$ & S  \\ 
2459102.3310 & $-54.3\pm2.1$ & $43.0\pm2.2$ & C  \\ 
2459102.6343 & $-104.4\pm2.6$ & $98.2\pm3.2$ & C  \\ 
2459102.6482 & $-107.1\pm2.2$ & $98.8\pm2.8$ & C  \\ 
2459103.3466 & $-92.5\pm1.7$ & $72.0\pm2.0$ & C  \\ 
2459103.3712 & $-89.6\pm1.9$ & $69.4\pm2.3$ & C  \\ 
2459103.6092 & $-68.4\pm2.2$ & $51.7\pm2.2$ & C  \\ 
2459114.3461 & $-107.7\pm1.3$ & $89.9\pm1.7$ & C  \\ 
2459114.5400 & $-83.6\pm3.2$ & $92.0\pm3.9$ & C  \\ 
2459114.5535 & $-92.0\pm4.8$ & $78.1\pm5.8$ & C  \\ 
2459116.3337 & $90.3\pm2.1$ & $-103.4\pm2.5$ & C  \\ 
2459116.3609 & $100.1\pm4.1$ & $-110.6\pm3.8$ & C  \\ 
2459121.3176 & $-76.2\pm2.5$ & $66.3\pm2.4$ & C  \\ 
2459121.4795 & $-101.3\pm1.7$ & $86.5\pm1.9$ & C  \\ 
2459122.3248 & $-86.7\pm2.7$ & $71.1\pm2.5$ & C  \\ 
2459122.5446 & $-63.4\pm2.8$ & $61.5\pm2.7$ & C  \\ 
2459122.5625 & $-65.3\pm3.0$ & $58.9\pm2.7$ & C  \\ 
2459128.2973 & $138.4\pm2.3$ & $-144.6\pm2.6$ & C  \\ 
2459133.3468 & $-100.4\pm2.2$ & $89.7\pm2.4$ & C  \\ 
2459133.5187 & $-92.3\pm2.6$ & $82.7\pm3.0$ & C  \\ 
2459135.3393 & $107.8\pm1.8$ & $-111.0\pm1.7$ & C  \\ 
2459135.3726 & $106.5\pm1.6$ & $-115.5\pm1.4$ & C  \\ 
2459135.5200 & $132.5\pm1.8$ & $-135.6\pm2.1$ & C  \\ 
2459146.3043 & $49.2\pm7.3$ & $-48.8\pm7.0$ & C  \\ 
2459146.4732 & $85.6\pm4.9$ & $-82.9\pm4.9$ & C  \\ 
2459147.4269 & $61.7\pm3.5$ & $-67.6\pm4.3$ & C  \\ 
2459148.3252 & $-96.7\pm5.8$ & $96.5\pm5.5$ & C  \\ 
2459148.4267 & $-106.5\pm2.9$ & $98.5\pm3.1$ & C  \\ 
\noalign{{\smallskip}}
\hline
\end{tabular}
\end{table}

\end{appendix}
\end{document}